\documentclass[fleqn,usenatbib]{mnras}

\usepackage[T1]{fontenc}

\usepackage{graphicx}	
\usepackage{amsmath}	
\usepackage{amssymb}	

\usepackage{mathtools}
\usepackage{diagbox}
\usepackage{multirow, makecell}
\usepackage{multicol}
\usepackage[flushleft]{threeparttable}
\newcommand\Tstrut{\rule{0pt}{2.6ex}}         
\usepackage{newtxtext,newtxmath}

\title[Atmospheric parameters of M Dwarfs]{M dwarf spectral indices at moderate resolution: accurate T$_\text{eff}$ and [Fe/H] for 178 southern stars\thanks{Based on observations collected at Observatório do Pico dos Dias (OPD), operated by the Laboratório Nacional de Astrofísica, MCTI, Brazil.}}

\author[Costa-Almeida et al.]{
Ellen Costa-Almeida,$^{1,2}$\thanks{Present Address: Observatório Nacional/MCTI, Rio de Janeiro, RJ, Brazil. Email Addresses: ellenalmeida@on.br (ECA); gustavo@astro.ufrj.br (GFPdM).}
Gustavo F. Porto de Mello,$^{1}$
Riano E. Giribaldi,$^{3}$
Diego \newauthor \ Lorenzo-Oliveira,$^{4}$
Maria L. Ubaldo-Melo$^{1}$
\\
$^{1}$Observatório do Valongo, Universidade Federal do Rio de Janeiro, Ladeira do Pedro Antônio 43, 20080-090 Rio de Janeiro, RJ, Brazil\\
$^{2}$Observatório Nacional/MCTI, Rua General José Cristino 77, 20921-400 Rio de Janeiro, RJ, Brazil\\
$^{3}$Nicolaus Copernicus Astronomical Center, Polish Academy of Sciences, Bartycka 18, 00-716, Warsaw, Poland\\
$^{4}$Universidade de São Paulo, Departamento de Astronomia do IAG/USP, Rua do Matão 1226, Cidade Universitária, 05508-900 São Paulo, SP, Brazil
}

\date{Accepted 2021 September 28. Received 2021 September 28; in original form 2021 May 11}

\pubyear{2020}

\begin{document}
\label{firstpage}
\pagerange{\pageref{firstpage}--\pageref{lastpage}}
\maketitle

\begin{abstract}
We present a spectroscopic and photometric calibration to derive effective temperatures T$_\text{eff}$ and metallicities [Fe/H] for M dwarfs, based on a Principal Component Analysis of 147 spectral indices measured off moderate resolution (R$\sim$11,000), high S/N ($>$100) spectra in the $\lambda \lambda$ 8390-8834 region, plus the J$-$H color. Internal uncertainties, estimated by the residuals, are 81 K and 0.12 dex, respectively, for T$_\text{eff}$ and [Fe/H], the calibrations being valid for 3050 K $<$ T$_\text{eff}$ $<$ 4100 K and $-$0.45 $<$ [Fe/H] $<$ $+$0.50 dex. The PCA calibration is a competitive model-independent method to derive T$_\text{eff}$ and $[$Fe/H$]$ for large samples of M dwarfs, well suited to the available database of far-red spectra. The median uncertainties are 105 K and 0.23 dex for T$_\text{eff}$ and [Fe/H], respectively, estimated by Monte Carlo simulations. We compare our values to other works based on photometric and spectroscopic techniques and find median differences 75 $\pm$ 273 K and 0.02 $\pm$ 0.31 dex for T$_\text{eff}$ and [Fe/H], respectively, achieving good accuracy but relatively low precision. We find considerable disagreement in the literature between atmospheric parameters for stars in common. We use the new calibration to derive T$_\text{eff}$ and [Fe/H] for 178 K7-M5 dwarfs, many previously unstudied. Our metallicity distribution function for nearby M dwarfs peaks at [Fe/H]$\sim$-0.10 dex, in good agreement with the RAVE distribution for GK dwarfs. We present radial velocities (internal precision 1.4 km/s) for 99 objects without previous measurements. The kinematics of the sample shows it to be fully dominated by thin/thick disk stars, excepting the well-known high-velocity Kapteyn’s star.

\end{abstract}

\begin{keywords}
stars: fundamental parameters -- stars: low-mass -- techniques: spectroscopic -- stars: abundances -- Galaxy: solar neighborhood
\end{keywords}



\section{Introduction}

M dwarfs represent more than 70\% in number and about 40\% of the stellar mass content of the Galaxy (\citealt{Kirkpatrick2012}), ranging from 0.08 M$_{\odot}$ to 0.6 M$_{\odot}$ -- even though an extension towards even less massive objects is possible if they are very young, see \cite{Sahlmann2021} -- and, along with the fact that they have main-sequence lifetimes exceeding the Hubble time, they are fundamental assets towards a realistic understanding of the formation, structure and chemodynamical evolution of the Galaxy. Because of their very high number density, they are attractive targets in the search for exoplanets, especially rocky planets, since the sensitivity of the transit and radial-velocity techniques performs better for higher planet-to-star mass and radius ratios. Estimates show that there are at least 3 planets per M dwarf (\citealt{Tuomi2019}) and that smaller planets are more likely to be formed around smaller stars (e.g., \citealt{Bonfils2013}; \citealt{DC2013}; \citealt{Tuomi2014}). Therefore, there is an urge for accurate, extensive and homogeneous stellar properties of M dwarf stars.

The precise estimation of atmospheric parameters of M dwarfs is a theoretical and observational challenge, especially when compared to the classical model atmospheres methods applied for solar-type stars, which are very precise and accurate. M dwarfs are so cool that diatomic and triatomic molecules can survive in their atmospheres, filling the spectra with many absorption bands and making it very difficult, and often impossible, to identify the continuum itself and individual absorption lines. Because of this, most of the techniques used to derive atmospheric parameters, particularly effective temperature and metallicity (hereafter, T$_\text{eff}$ and [Fe/H], respectively), are based on calibrations using photometric colors (e.g., \citealt{Bonfils2005}, \citealt{Casagrande2008}, \citealt{Neves2012}) or spectral indices (e.g., \citealt{Rojas-Ayala2012}, \citealt{Mann2013a}, \citealt{Neves2014}, \citealt{Newton2015}, \citealt{Lopez-Valdivia2019}).

Regarding T$_{\text{eff}}$, interferometry has been used to derive \textit{practically model-independent} precise estimates for bright and nearby M dwarfs with uncertainties smaller than 100 K by using relationships between stellar radius and T$_{\text{eff}}$ (e.g., \citealt{Boyajian2012b}; \citealt{vonBraun2014}). However, there is only a limited number of stars with radii measured by this technique, especially in the southern hemisphere (e.g., \citealt{Segransan2003}; \citealt{Rabus2019}). When we consider differences between recent T$_{\text{eff}}$ determinations, we find offsets reaching up to 300 K between model-dependent studies. Our lack of comprehension of interiors and atmospheres of almost fully convective low-mass stars, together with our incomplete database of atomic and molecular transitions, are mainly responsible for the discrepancies observed in the literature. Therefore, studies that use interferometric T$_{\text{eff}}$ estimations as calibrators offer advantages in accuracy and should be preferred for the systematic derivation of atmospheric parameters from indirect methods. We have recently showed the competitiveness of line indexes to derive accurate and precise atmospheric parameters for solar-type stars (e.g., \citealt{Luan2014}, \citealt{Giribaldi2019}) as long as a suitable sample of calibrating stars is employed and systematic errors are carefully checked and accounted for. Additional efforts to extend this technique to lower T$_{\text{eff}}$ domain, thus, appear to be fully warranted.


Regarding [Fe/H]\footnote{Hereafter defined as [Fe/H]$_\star$ = $\mathrm{log} \left(\frac{N_\mathrm{Fe}}{N_\mathrm{H}}\right)_\star-\ \mathrm{log} \left(\frac{N_\mathrm{Fe}}{N_\mathrm{H}}\right)_\odot$, where $N_\mathrm{Fe}$ and $N_\mathrm{H}$ are the number density of Fe and H atoms, respectively.}, a common practice is to employ wide physical binaries systems of FGK + M dwarf by assuming that the composition of the hotter star, which usually has a precise [Fe/H] determination with typical uncertainties around 0.07 dex, can be extrapolated to its cooler companion (e.g., \citealt{Neves2012}; \citealt{Mann2013b}, \citeyear{Mann2015}; \citealt{Montes2018}), although considering the high frequency of planets around M dwarfs, planet accretion scenarios can change the original composition of the star and should be considered with caution (e.g., \citealt{Teske2015}; \citealt{Oh2018}). Moreover, typical uncertainties around 0.10 dex are presently achievable using spectral features in the NIR (e.g., \citealt{Rojas-Ayala2012}; \citealt{Mann2013a}; \citealt{Neves2014}) and around 0.20 dex using photometric calibrations (e.g., \citealt{Bonfils2005}; \citealt{Neves2012}; \citealt{Newton2014}), even though the agreement of values between methods is less than perfect. We refer the reader to \cite{Lindgren2016} who reported differences of $\sim$0.6 dex for individual stars comparing different calibrations.

Accurate [Fe/H] determinations are essential for a better understanding of the composition of exoplanets. Nowadays, it is widely established that, on average, metal-rich FGK stars have a higher frequency of planets (e.g., \citealt{Santos2004}; \citealt{VF05}; \citealt{Ghezzi2018}). However, recent studies are finding evidence that this trend is also true for host M dwarfs (e.g., \citealt{Hobson2018}). Thus, the previous knowledge of [Fe/H] is important to focus the search for interesting planet-hosting M dwarfs for future direct observations of their planets with upcoming instrumentation such as GMT, JWST and LUVOIR.

In our work we present accurate model-independent T$_{\text{eff}}$ and [Fe/H] determinations for 178 southern M dwarf stars in the solar neighbourhood. We explore the equivalent widths (hereafter, EW) of 147 spectral indices in the far red together with $J-H$ color to derive the Principal Components of the Principal Component Analysis (hereafter, PCA). Our atmospheric parameter scale is calibrated with 44 stars from \citealt{Mann2015}, who used more than 20 stars with interferometric T$_{\text{eff}}$ determinations as calibrators. In Sect. \ref{sec2} we describe the sample selection and define the calibration stars. In Sect. \ref{sec3} we describe the characteristics of the spectroscopic observations of the 178 program stars, the measurement of radial velocities, the calculation of the stellar space velocities and Galactic orbits, and the procedure for normalizing the continuum.
In Sect. \ref{sec4} we describe the line index system, its definition criteria and associated errors. In Sect. \ref{sec5} we present the calibrations for T$_{\text{eff}}$ and [Fe/H]. In Sect. \ref{sec6} we describe the Monte Carlo simulations performed to gauge the errors of the calibrations and present the final atmospheric parameters. We also compare our results to previously published values derived using other techniques. In Sect. \ref{sec:dynamics} we analyse the Galactic orbits of the M dwarfs and associate them to the Galactic substructures. In Sect. \ref{sec7} we summarize our conclusions.
\section{Sample Selection}
\label{sec2}
This project started on 2007, as a pilot, initially targeting a sample of 94 stars selected from the Hipparcos Catalog (\citealt{Hipparcos97}) which, by that time, had the most precise parallaxes available. The sample is volume limited ($\pi > 100$ mas, i.e., $d<10$ pc) and emphasized stars brighter than $V\sim11$ mag. We observed this initial sample on 5 observing runs, from 2007 to 2011. The observations were resumed in 2016 and the sample was considerably expanded by adding stars from \cite{Winters2015} having trigonometric distances $d\leq$ 20 pc and $V\leq$13 mag. By the end of 6 additional observing runs from 2017 to 2018, we observed in total 178 southern ($\delta \lesssim +20$) K7V-M5.5V stars.

\begin{figure}
	\includegraphics[width=\columnwidth]{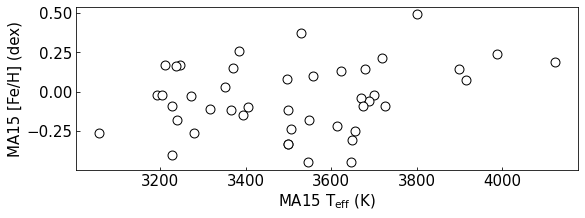}
    \caption{Distribution of the atmospheric parameters of the calibration stars. Values taken from \citealt{Mann2015}.}
    \label{fig:param_calib_stars}
\end{figure}

We completed our observation runs having 44 stars in common with \cite{Mann2015} (hereafter, MA15), who derived T$_{\text{eff}}$ by comparing optical spectra with PHOENIX atmosphere models and obtained [Fe/H] from EWs of Ca and Na atomic lines in the NIR (\citealt{Mann2013a} and \citealt{Mann2014}). These objects are henceforth referred to as ``calibration stars'' and the atmospheric parameter space they cover is shown in Figure \ref{fig:param_calib_stars}, 3056 to 4124 K and -0.45 to +0.49 dex for T$_{\text{eff}}$ and [Fe/H], respectively. Although our calibration sample has a bias for stars with T$_{\text{eff}}>$ 3800 K, in the sense that hotter stars are all metal-rich, we have kept these K stars in common with MA15 to investigate if the spectral index method would be amenable to extrapolation.




\section{Observations and reduction}
\label{sec3}
\subsection{Coud\'{e} spectroscopy}
We used the coud\'e spectrograph fed by the 1.60m Perkin-Elmer telescope of Observat\'orio do Pico dos Dias (OPD, Braz\'opolis, Brazil), operated by Laborat\'orio Nacional de Astrof\'isica (LNA/CNPq). The old (2007 to 2012) and new (2017 to 2018) data were obtained using two different CCDs, yet possessing very similar specifications, both having 2048 pixels and 13.5 $\mu$m/pixel. The slit width was adjusted to 500 $\mu$m to give a 3-pixel resolving power R $\approx$ 11000 and a 600 l/mm diffraction grating was employed in the first order, obtaining 0.25 \AA/pixel. We used a yellow RG 610 filter to block contamination from the second order.  The spectral region is centered at 8650 \AA, with a spectral coverage of 500 \AA\ (8370 to 8870 \AA). One clear advantage of this range as compared to bluer regions is the much diminished line blanketing, making the local continuum much more accessible, at least for the hotter subtypes. This fact lends more credence to the spectral indices, which are probably reflecting true physical absorption more realistically. Our sample stars were sufficiently faint for exposure times to often be substantial. 

\begin{figure}
	\includegraphics[width=\columnwidth]{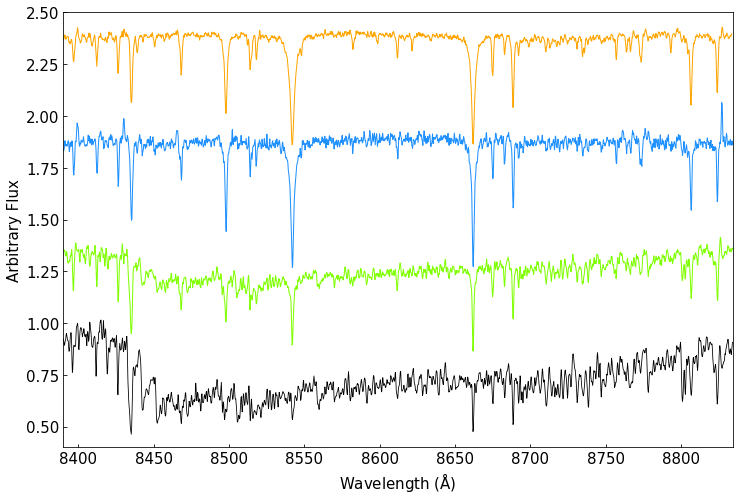}
    \caption{Sample spectra of hot, intermediate and cool M dwarfs for average quality spectra. From top to bottom we show HIP 40239 (M0V, orange line), HIP 72511 (M1.5V, blue line), HIP 53020 (M4V, green line) and HIP 70890 (M5.5V, black line) having nominal S/N of 190, 160, 150 and 180, respectively.}
    \label{fig:example_spectra}
\end{figure}

We divided the exposure times based on V magnitude values to achieve nominal S/N greater than 100 for all spectra (see Table \ref{tab:exptime}). These stars do not have clear continuum windows free of absorption lines to directly estimate the continuum fluctuations, so we estimated the nominal S/N ratios by means of Poissonian photon statistics probably slightly overestimating the actual S/N ratios. Sample spectra of cool, intermediate and hot M dwarfs are shown in Figure~\ref{fig:example_spectra}, for average quality spectra. The data reduction was carried out by standard techniques with a Python script calling IRAF\footnote{Image Reduction and Analysis Facility (IRAF) is distributed by the National Optical Astronomical Observatories (NOAO), which is operated by the Association of Universities for Research in Astronomy (AURA), Inc., under contract to the National Science Foundation (NSF).} tasks through Pyraf.

\begin{table}
    \centering
    \caption{Average exposure times and approximate nominal S/N ratios for the averaged spectra.}
    \begin{tabular}{ccc}
        \hline
        \hline
        $V$ & Exposure time & Nominal S/R\\
        (mag) & (s) & \\
        \hline
        $V \leq 10.0$ & $3 \times 600$ & 150\\
        $10.0 < V \leq 10.5$ & $3 \times 900$ & 150\\
        $10.5 < V \leq 11.5$ & $4 \times 900$ & 150\\
        $11.5 < V \leq 12.0$ & $4 \times 900$ & 150\\
        $12.0 < V \leq 12.5$ & $5 \times 900$ & 100\\
        $12.5 < V \leq 13.0$ & $6 \times 1200$ & 100\\
        \hline
    \end{tabular}
\label{tab:exptime}
\end{table}

\subsection{Radial velocities and Galactic kinematic parameters}
Doppler shift correction is a standard procedure in data reduction, but since these stars were poorly represented in RV catalogs up until Gaia Data Release 2 (\citealt{GaiaDR2}, Gaia DR2), we are presenting our values of radial velocity. 

We selected 10 relatively isolated spectral lines spread along the full spectral coverage and used the central wavelengths values at rest from the version 5.7 of the NIST Standard Reference Database 78\footnote{\url{https://www.nist.gov/pml/atomic-spectra-database}} (see Table \ref{tab:linhas_dopcor}). For each observing run, we selected a hot (K, M0 or M1) and a cool (M2, M3 or M4) star to be our template spectra and manually measured the observed central wavelengths ($\lambda_{\mathrm{obs}}$) of the 10 lines using the \textit{splot} IRAF task. We used the mean value and the standard deviation as final observed velocity and associated error, respectively. We used the task \textit{dopcor} for the Doppler shift correction and obtained spectra at rest (template spectra). 

\begin{table}
    \centering
    \caption{Atomic lines used for the Doppler correction.}
    \begin{tabular}{cc}
        \hline
        \hline
        $\lambda_\text{central}$  & Element\\
        \hline
        8468.407 & Fe I \\
        8611.803 & Fe I \\
        8621.600 & Fe I \\
        8633.956 & Ca I \\
        8682.987 & Ti I \\
        8688.625 & Fe I \\
        8757.187 & Fe I \\
        8763.966 & Fe I \\
        8806.575 & Mg I \\
        8824.221 & Fe I \\
        \hline
    \end{tabular}
\label{tab:linhas_dopcor}
\end{table}

\begin{figure}
	\includegraphics[width=\columnwidth]{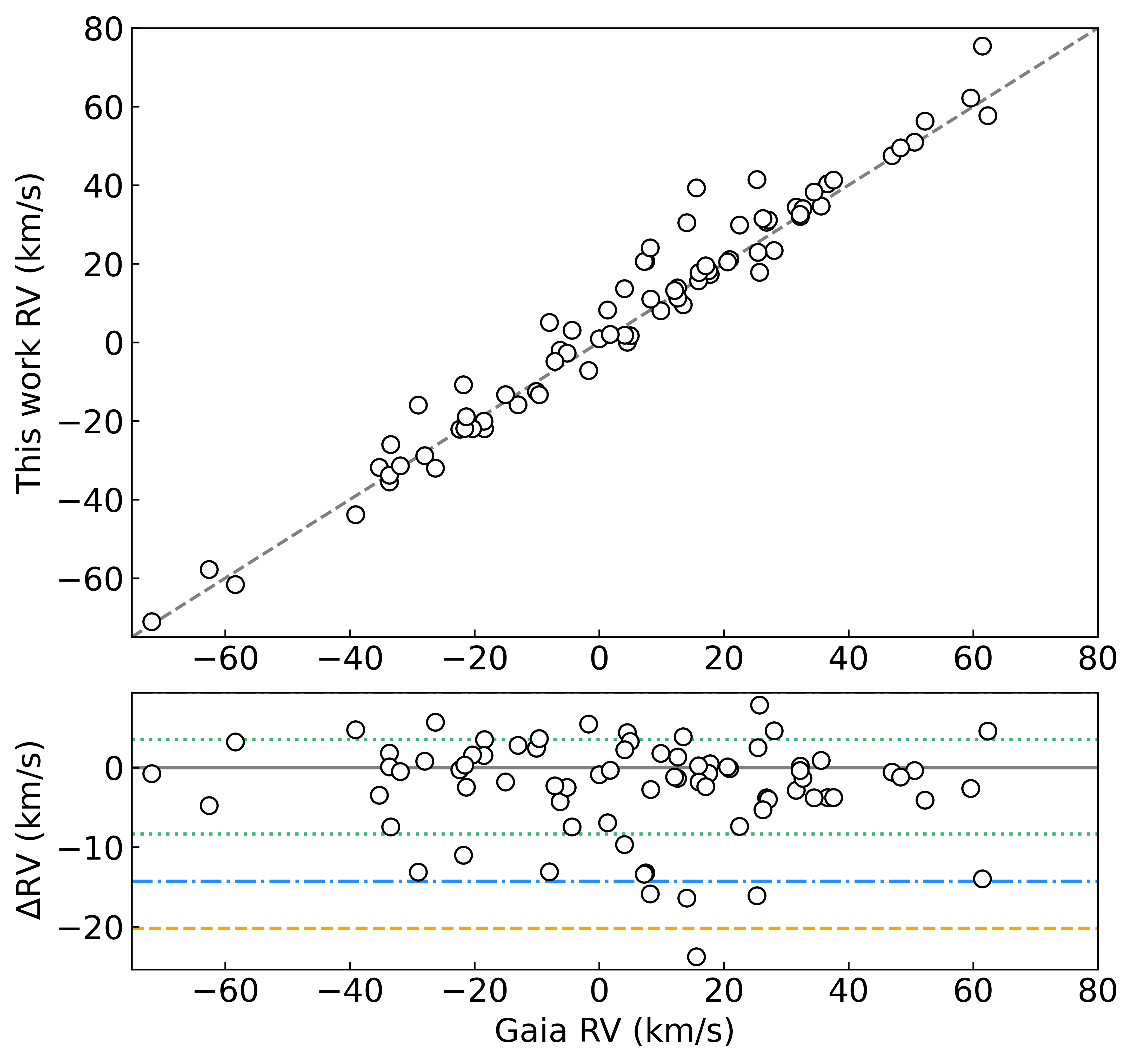}
    \caption{Upper pannel: Comparison between radial velocities from this work for 79 stars in common with measured RV velocities on Gaia Data Release 2. The dashed grey line represents equality.
    Bottom panel: RV comparison (Gaia-this work) between Gaia and this work. The differences have a RMS scatter of 5.92 km/s and a zero-point difference of -2.42 km/s. For this sample, the median errors of Gaia radial velocities is 0.245 km/s and ours is 1.243 km/s. The grey solid line represents the zero, green dotted lines are $\pm$1$\sigma$, blue dash-dotted lines are $\pm$2$\sigma$ and orange dashed line represents 3$\sigma$ of the distribution. The four high residuals ($>2\sigma$) are identified as HIP67155, HIP 439, HIP 65859 and HIP 114046 having radial velocities of 39.3$\pm$1.3, 41.1$\pm$1.4, 30.5$\pm$1.2 and 34.1$\pm$1.5 km/s, respectively. None of them are known binary systems, and we note that HIP 114046 has two confirmed planets (\citealt{Jeffers2020}).}
    \label{fig:rv_gaia}
\end{figure}

Considering that molecular absorption changes dramatically for different M spectral subtypes, turning a M4V spectrum really different from a M0V, using hot/cool template spectra was extremely important for performing a Fourier cross-correlation on the input list of objects and template spectra.  This procedure produced for the whole sample relative and heliocentric velocities and the associated error.  The overall internal precision, estimated from the median of our velocity uncertainties, is 1.40 km/s. 
We compared our results with Gaia Data Release 2 radial velocities and found a mean difference of $-2.42\pm5.92$ km/s for 79 stars in common (see Figure \ref{fig:rv_gaia}). A possible slight underestimation of our RV values with respect to Gaia DR2 is suggested, but without clear statistical significance regarding the internal errors of both. We thus provide new RVs for 99 stars without any previous data from Gaia and the RV results are given in Table \ref{tab:PCA}.

We integrated stellar orbits running the code GalPot\footnote{\url{https://github.com/PaulMcMillan-Astro/GalPot}} \citep{McMillan2017} with which we obtained the integrals of motion: orbital energy and the angular momentum component perpendicular to the Galactic plane $L_{Z}$. 
For the calculations, we adopted the best-fit Galactic potential model of \citep{McMillan2017} and used the Galactic space-velocity components ($R_{\mathrm{Gal}}, \phi, Z-U, V_\phi, W$) derived from the radial velocities in  Table~\ref{tab:PCA}, and the coordinates, parallaxes, and proper motions from the Gaia EDR3 catalog \citep{GaiaEDR3_2021}.
The frame transformations were performed using Astropy \citep{astropy2013A&A...558A..33A, astropy:2018} and PyAstronomy routines \citep{pya} adopting the Galactocentric solar position in \citet{McMillan2017} and the solar velocity respect to the local standard of rest determined by \citet{2010MNRAS.403.1829S}.
The parameters above will be used in Sect.~\ref{sec:dynamics} to analyse the dynamic distribution of the M dwarfs in the Galactic substructures.
For comparison purposes, the same procedure was also applied to $5\,000$ random stars from the Gaia DR2 catalog restricted to $\pi >$ 10~mas with precisions better than 10\% ($\sigma_\pi/\pi$ < 0.1) and available radial velocities.

\subsection{Normalization}
We divided sample stars into hot (K dwarfs, M0V and M1V) and cool (higher than M2V) groups. The pseudo continuum was well defined for the hot group and thus a low order polynomial function sufficed to establish the local continuum. The absorption from molecules, specially TiO bands, is responsible for a substantial depression in the pseudo continuum of cooler stars. We therefore, after some tests, simply connected the two highest local continuum points with a straight line (usually 8400 \AA\ and 8850 \AA), considering these as the only local continuum points (see Figure \ref{fig:example_spectra}).
\section{Line index definition and measurement}
\label{sec4}
Spectral index is a region of the spectrum which contains a group of atomic and/or molecular lines that can not be individually distinguished but, as a group, are sensible to the variation of one or more atmospheric parameters. Its definition is made based on a visual inspection of the spectrum, i. e., there is no need to know about the physical properties of the star beforehand. Furthermore, this technique does not depend on models of internal structure or stellar atmosphere, turning it directly applicable to the spectrum without the necessity to stipulate hypothesis nor make previous interpretations in respect to the definition of the indices. 



\subsection{Definition of the indices}
\begin{figure*}
    \centering
    \resizebox{16cm}{!}{\includegraphics{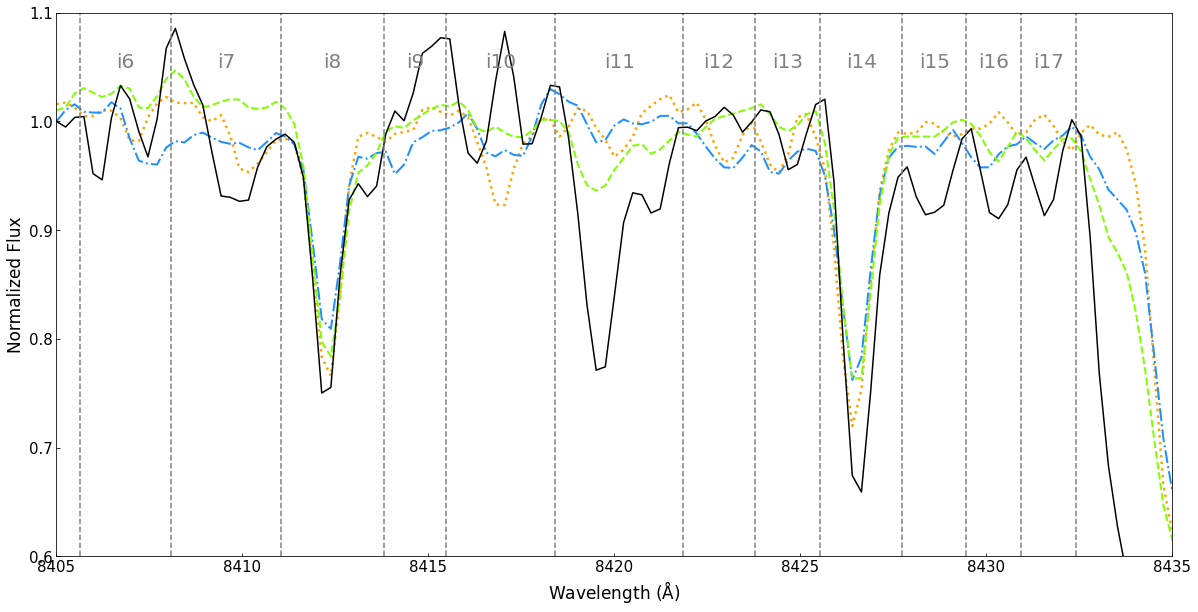}}
    \caption{Example of the definition of the spectral indices. Each index is represented as $i$ followed by a number (from 1 to 170) for increasing central wavelength. The dashed vertical grey lines represent the initial and final limits of each index. HIP 99701 (M0V) is the orange dotted line, HIP 51317 (M2V) is the blue dash-dotted line, HIP 87937 (M4V) is the green dashed line and Proxima Centauri (M5.5V) is the black solid line.}
    \label{fig:example_indices}
\end{figure*}

The definition of the spectral indices was made based on a visual inspection of the spectra of 4 representative stars of spectral subtypes that compose our sample: M0V, M2V, M4V and M5.5V. We over-plotted these spectra, looked for regions with a clear similar variation of the flux in the four of them and used the initial and final wavelengths of the coolest star to stipulate the limits to each group of lines (see Figure \ref{fig:example_indices}). We used all the spectrum coverage to define the indices, i. e., no region was ignored and there is no region in common between two indices. We defined 170 indices from 8390.17 to 8834.47 \AA\ (see Table \ref{tab:appendix_indices}).

The spectral indices in this region are dominated by neutral iron-peak species (such as 14 Fe I, 19 Ti I and 1 Mg I lines), ionized species (such as 3 Ca II lines) and molecular bands (such as 2 TiO, 1 FeH and 1 VO bands). For a detailed description of the spectral lines in this spectral range we refer the reader to \cite{Carmenes}. Besides, taking into consideration the possible chromospheric effects in the spectrum caused by the magnetic activity of the star, we did not use the indices constituted by the Ca II triplet lines (\textit{i}42, \textit{i}60 and \textit{i}105) because these lines can have a meaningful chromospheric filling-in (e.g., \citealt{Lorenzo-Oliveira2016}), thereby polluting any correlation with T$_\text{eff}$ and [Fe/H].

\subsection{Sanity check and associated errors}
\begin{figure}
	\includegraphics[width=\columnwidth]{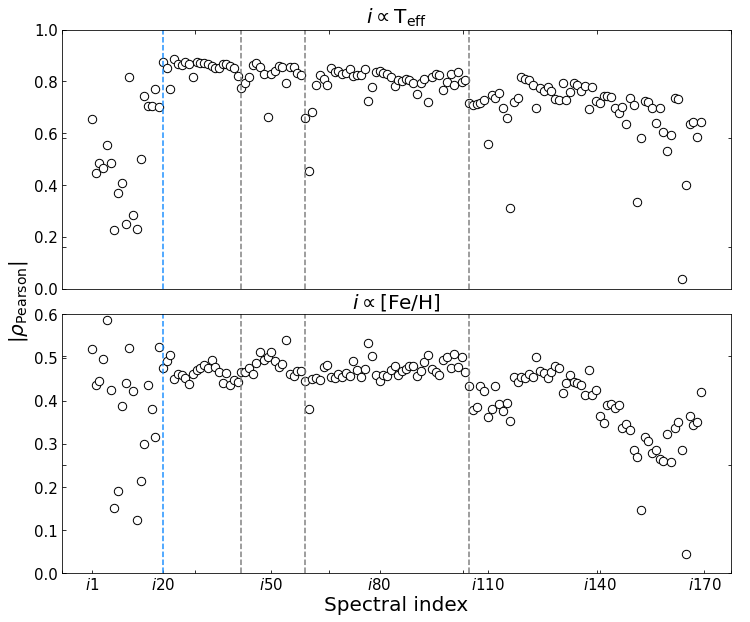}
    \caption{Module of the Pearson correlation coefficient for each index with respect to T$_{\text{eff}}$ (upper panel) and [Fe/H] (lower panel). The blue dotted line indicates the 20th index (see text) and the 3 grey dashed lines indicate the Ca II triplet lines.}
    \label{fig:pearson_indices}
\end{figure}

We verified if the indices had an appropriate physical behavior to decide which ones would be used in the next step. We certified the sensitivity of each index with respect to both T$_\text{eff}$ and [Fe/H] for the calibration sample by estimating the Pearson correlation coefficient (see Figure \ref{fig:pearson_indices}). The correlation visibly reduces as the wavelength increases, except for the initial part of the spectra which doesn't present a clear pattern on both graphics. This may be due to the presence of strong telluric absorption lines in this region (e.g., \citealt{Matheson2000}) for stars observed (9 calibration stars have more than one observation) on nights in which humidity was particularly high. In light of this, we decided not to use the first 19 indices for our analysis.





The consistency and repeatability of the EWs is extremely important for our method. The percent variation of the EW measurement will represent the associated error for each index. Between all observed stars, 41 had more than one spectrum. We calculated the relative error of each index of each star 
($\sigma^{\text{rel}}_{\star}$)
\begin{equation}
\label{relative_error_star}
\centering
    \sigma^{\text{rel}}_{\star} = \quad\begin{bmatrix}
    \frac{\sigma_{\text{EW}_{i20}}}{\text{med}(\text{EW}_{i20})} & \hdots & \frac{\sigma_{\text{EW}_{i170}}}{\text{med}(\text{EW}_{i170})} \end{bmatrix} \text{.}
\end{equation}
We obtained a distribution of 41 relative errors for each one of the 147 spectral indices. We assume that the final relative error is the median of the distribution of each index
\begin{equation}
\label{repetibilidade:eq4}
    \centering
    \sigma^{\text{rel}} = \quad\begin{bmatrix}
    \text{med}(\sigma^{\text{rel}}_{i20}) & \hdots & \text{med}(\sigma^{\text{rel}}_{i170})\end{bmatrix} \text{.}
\end{equation}

We found that the EWs are stable with a mean variation of 11$\pm$8 \% (1$\sigma$). Only $i149$ and $i157$ show relative errors greater than the 3$\sigma$ cut, respectively 49\% and 47\%. We estimate how these relative errors can modify our results in the next Section, but, for now, we remark that the final contribution of each index will be diluted within the Principal Components, reducing their particular impact. The initial and final wavelengths and final relative uncertainties of all the spectral indices can be found in Table \ref{tab:appendix_indices}.
\section{Methods}
\label{sec5}

Towards the goal of reducing the total number of variables to work with, turning them independent from one another and not over fitting our data, we use the Principal Component Analysis (PCA), which is a technique of feature extraction. In addition, it should be noted that PCA is a model-independent mathematical tool, but since it finds the best components to replicate the trend observed in the calibration sample atmospheric parameters, it also replicates its offsets and their model dependence. Our work follows very closely the methods exposed by \cite{Giribaldi2019} and thus a full discussion is unwarranted. A short description of the variables used to create the PCs and the regression strategy is given below.

\subsection{Incorporating 2MASS photometry}
The initial approach was to create the PCs based exclusively on the 147 spectral indices -- a purely spectroscopic approach. We found a great correlation between T$_{\text{eff}}$ and the first 3 PCs, i.e., the components that exhibit the 3 greatest variances of our data. This result was expected since T$_{\text{eff}}$ determines most of the spectrum shape. However, the [Fe/H] presented a poor correlation between the first 7 PCs -- which were responsible for 98.7\% of the total cumulative variance of the data--, so we investigated which colors had great correlations with [Fe/H] by using the Pearson coefficient as a diagnostic. All stars observed have Gaia and 2MASS magnitudes, so we used colors having $G$, $G_{\text{RP}}$, $J$, $H$ and $K$ (see Table \ref{tab:colors}). 

\begin{table}
    \centering
    \caption{Pearson correlation between colors and atmospheric parameters.}
    \begin{tabular}{lcc}
        \hline
        \hline
        Color & $\rho_{\text{T}_{\text{eff}}}$ & $\rho_{\text{[Fe/H]}}$\Tstrut\\[0.08cm]
        \hline
        $G-J$ & -0.96 & 0.015 \\
        $G-H$ & -0.93 & 0.124 \\
        $G-K$ & -0.93 & 0.106 \\
        $G_{\text{RP}}-J$ & -0.96 & 0.003 \\
        $G_{\text{RP}}-H$ & -0.95 & 0.06 \\
        $G_{\text{RP}}-K$ & -0.95 & 0.06 \\
        $J-H$ & 0.51 & 0.77 \\
        $J-K$ & 0.03 & 0.63 \\
        $H-K$ & -0.59 & -0.09 \\
        \hline
    \end{tabular}
\label{tab:colors}
\end{table}

It is clear from Table \ref{tab:colors} that $J-H$ and $J-K$ have the best correlations with [Fe/H]. Adding to that, we ran some tests to decide which and how many colors to use besides the 147 spectral indices. We found that it is not worth including colors to improve the correlation between T$_{\text{eff}}$ and the PCs because the spectral indices have more than enough predictive power by themselves. By including them, we are only overdetermining T$_{\text{eff}}$ and increasing the initial number of variables. Moreover, $J-H$ and $J-K$ together do not significantly change the correlations between [Fe/H] and the PCs, $J-H$ being enough. Considering this, we created our principal components based on $J-H$ color and 147 spectral indices from 44 calibration stars -- all having `A' 2MASS quality flag for both $J$ and $H$ magnitudes. Ten calibrations stars had more than one spectrum (57 spectra in total) but since adding variables with the same atmospheric parameters can create a bias around these values, we decided to use the mean EW of the indices of these stars. The color and spectral indices were standardized to take into account their different scales. The median and standard deviations of each variable used in the standardization procedure and the coefficients of the Principal Components created are listed in Table \ref{tab:PCA}.

\subsection{Calibrations by PCA}
\label{sec:calibrations_by_PCA}
We explored the correlations between the first 5 principal components -- which accounts for 97.8\% of the total cumulative variance of the data -- and the atmospheric parameters. The other higher principal components were discarded because they do not show significant correlations.  T$_{\text{eff}}$ has great correlations with PC1, PC2 and PC3 while [Fe/H] is better described by PC3 and PC5. Although PC1 saturates approximately at 3800 K, in the same limit, PC2 and PC3 start to show a trend with effective temperature, helping to differentiate hotter stars. We used the best regressive model, i.e., smallest regressive error and greater degrees of freedom, to build calibrations for the atmospheric parameters as functions of the PCs (Equations \ref{calib:teff} and \ref{calib:feh}).

\begin{equation}
\label{calib:teff}
\centering
\begin{aligned}
    \text{T}_{\text{eff}} = 3509.49(\pm10.77) + 17.22(\pm0.94)\text{PC1} \\
    - 31.51(\pm3.99)\text{PC2} + 51.17(\pm7.09)\text{PC3} \text{;}    
\end{aligned}
\end{equation}

\begin{equation}
\label{calib:feh}
\centering
\begin{aligned}
    \text{[Fe/H]} = -0.04(\pm0.02) + 0.11(\pm0.01)\text{PC3} \\
    + 0.07(\pm0.02)\text{PC5} \text{.}    
\end{aligned}
\end{equation}

The internal uncertainties are 81 K and 0.115 dex and the Pearson correlations between the fitted and observes values are 0.96 and 0.82 for T$_{\text{eff}}$ and [Fe/H], respectively. Even though the correlation between [Fe/H] and PC5 is not visually clear, adding PC5 to the regression increases the Pearson coefficient by 0.05 (0.77 to 0.82) and reduces the dispersion of the fitted values by 0.0415 (0.1562 to 0.1148). The same goes for T$_{\text{eff}}$ where we find the Pearson coefficient increasing from 0.83 to 0.90 to 0.96, and the dispersion of the fitted values from 108.3 K to 99.49 K to 81.26 K for PC1, PC1+PC2 and PC1+PC2+PC3, respectively.

\begin{figure}
	\includegraphics[width=\columnwidth]{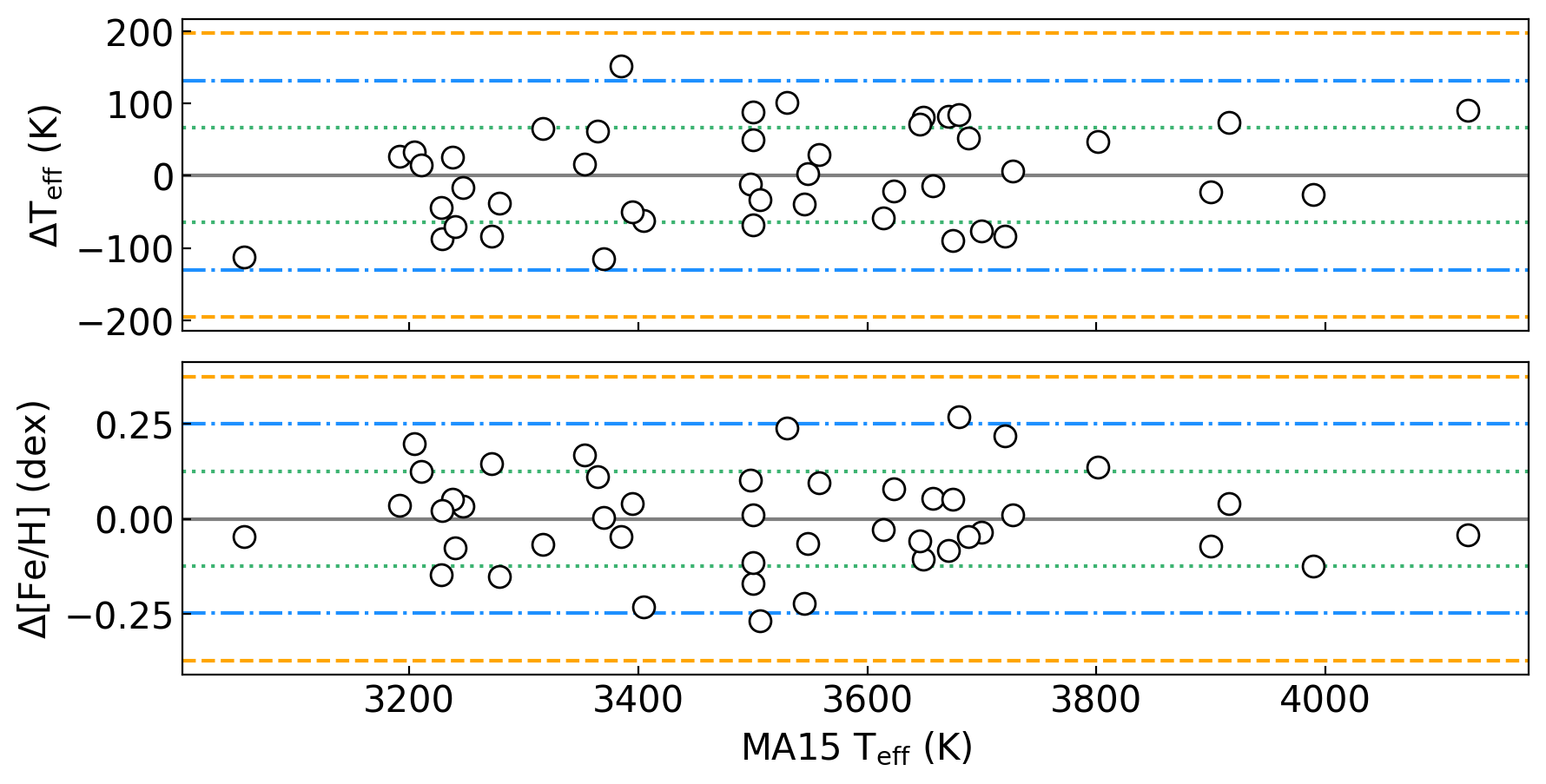}\\
	\includegraphics[width=\columnwidth]{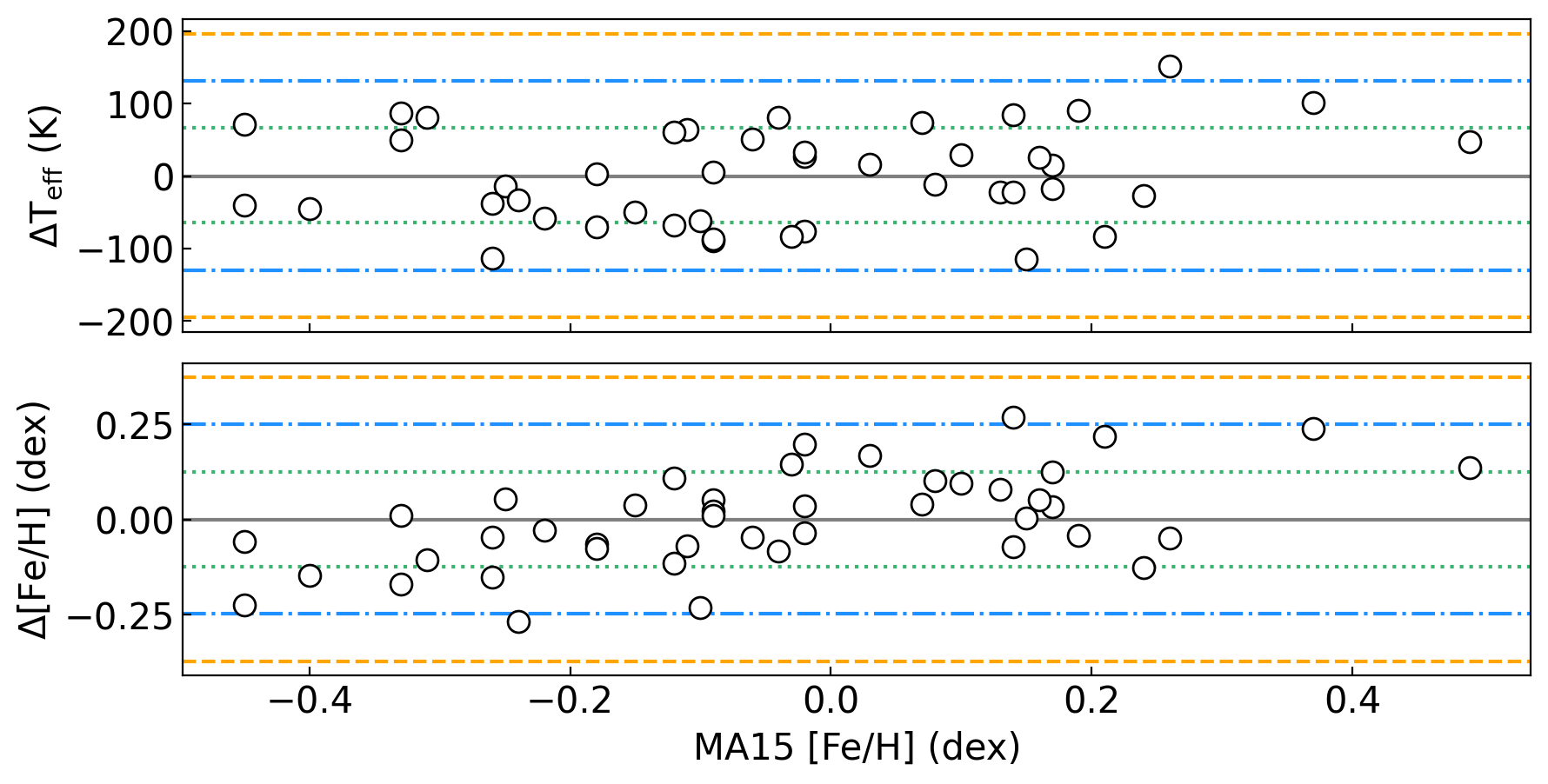}
    \caption{Residuals for both atmospheric parameters versus T$_{\text{eff}}$ and [Fe/H] from MA15 in the upper and bottom panels, respectively. T$_{\text{eff}}$ and [Fe/H] residuals have a RMS scatter of 65 K and 0.12 dex, respectively. Our residuals reach at most 2$\sigma$ approximately for both atmospheric parameters, finding the worst cases around the richer and hotter K dwarf stars. The grey solid lines represent the zero, green dotted lines are $\pm$1$\sigma$, blue dash-dotted lines are $\pm$2$\sigma$ and orange dashed lines are $\pm$3$\sigma$ of the distributions.}
    \label{fig:residuals_calibration}
\end{figure}

We analysed the trends between the residuals and the observed atmospheric parameters by running linear regressions (see Figure \ref{fig:residuals_calibration}). Table \ref{tab:regressions} shows that the only coefficient having a meaningful t-value (i.e., higher than 2) is the angular coefficient of the $\Delta{\text{[Fe/H]}}$ vs [Fe/H] regression. Considering our preferred simple calibration based on MA15, our residuals in [Fe/H] reach at most 0.25 dex and, taking into account these numbers, we chose to improve our [Fe/H] determinations only by applying a simple linear correction to our [Fe/H]$_{\text{PCA}}$ values, thus arriving at the corrected $\text{[Fe/H]}_{\text{PCA}}^{\text{corr}}$ values (Equation \ref{eq:corrected_feh}):

\begin{equation}
\label{eq:corrected_feh}
\centering
\begin{aligned}
    \text{[Fe/H]}_{\text{PCA}}^{\text{corr}} =  \frac{\text{[Fe/H]}_{\text{PCA}}}{0.6741(\pm 0.0762)} \text{.}    
\end{aligned}
\end{equation}



\begin{table}
    \centering
    \caption{Linear regressions between residuals and atmospheric parameters.}
    \begin{tabular}{lccc}
        \hline
        \hline
        Y (Residual) & X (Parameter) & $a_{0}$ (t-value) & $a_{1}$ (t-value) \\
        \hline
        $\Delta {\text{T}_{\text{eff}}}$ & $\text{T}_{\text{eff}}$ & -1.82 & 1.82 \\
        $\Delta {\text{T}_{\text{eff}}}$ & $\text{[Fe/H]}$ & 0.29 & 1.24 \\
        
        $\Delta {\text{[Fe/H]}}$ & $\text{T}_{\text{eff}}$ & 0.40 & -0.40 \\
        
        $\Delta {\text{[Fe/H]}}$ & $\text{[Fe/H]}$ & 1.03 & 4.28 \\
        \hline
    \end{tabular}
\label{tab:regressions}
\end{table}

\begin{figure}
	\includegraphics[width=\columnwidth]{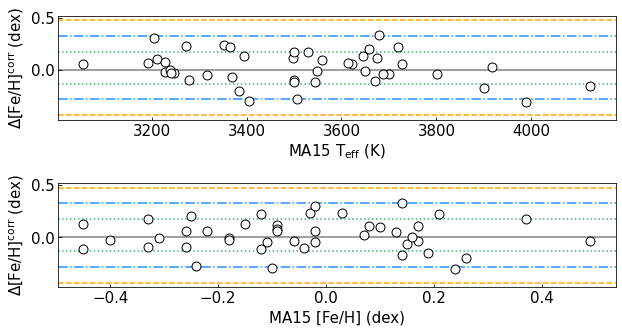}
    \caption{Corrected metallicity residuals versus observed T$_{\text{eff}}$ and [Fe/H] in the upper and bottom panels, respectively. We completely removed the trend with [Fe/H] and slightly decreased the trend with T$_{\text{eff}}$. The grey solid lines represent the zero, green dotted lines are $\pm$1$\sigma$, blue dash-dotted lines are $\pm$2$\sigma$ and orange dashed lines are $\pm$3$\sigma$ of the distributions.}
    \label{fig:corrected_residuals}
\end{figure}

We replotted the residuals with the new $\text{[Fe/H]}_{\text{PCA}}^{\text{corr}}$ values (see Figure \ref{fig:corrected_residuals}) and the trend with [Fe/H] was completely removed along with a slightly decrease of the trend with T$_{\text{eff}}$. We found -1.5 and 0 t-values for the angular coefficients of T$_{\text{eff}}$ and [Fe/H], respectively. The PCA method is heavily based on the line index strengths, and these are at least partially degenerate between the T$_{\text{eff}}$ and [Fe/H] values, so a degree of correlation between the results of these two parameters is expected. It is probably impossible, within the limitations of the technique we employ, to completely eliminate these trends, and this result speaks clearly of the difficulties in deriving consistently both $T_{\text{eff}}$ and [Fe/H] for red dwarfs when both parameters are determined still very far from fundamental considerations (i.e, hypothesis free), unlike what is nowadays possible for FGK stars (e.g., \citealt{gaia_benchmark}). These values are, in our judgment, the best T$_{\text{eff}}$ and [Fe/H] determinations from the line index measurements coupled to the $J-H$ color.

\section{Atmospheric parameters determination}
\label{sec6}
\subsection{Final effective temperature and [Fe/H] and associated errors}
\begin{figure}
	\includegraphics[width=\columnwidth]{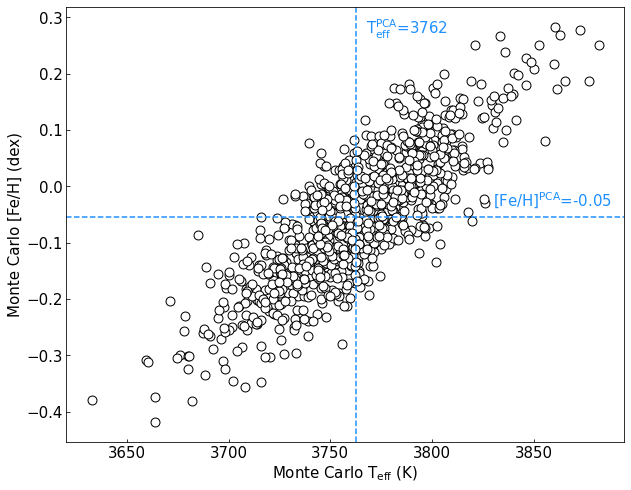}
    \caption{Distribution of the atmospheric parameters found for HIP 51007 (M1V). The dashed blue lines represent the median of each atmospheric parameter, i.e., 3763 K for $T_{\mathrm{eff}}^{\mathrm{PCA}}$ and -0.05 dex for [Fe/H]$^{\mathrm{PCA}}$.}
    \label{fig:MC_distribution}
\end{figure}

To properly explore the full uncertainty range in the EWs of the spectral indices, we performed 1000 Monte Carlo (MC) simulations assuming that the EWs errors follow Gaussian distributions ($J-H$ values were not simulated). 
Thus, each spectrum originated 1000 different spectra and, consequently, a distribution of atmospheric parameters from which the most probable value was located at the center of the distribution (see Figure \ref{fig:MC_distribution}). 
The final $T_{\mathrm{eff}}$ and [Fe/H] values found for each spectrum, hereafter called $T_{\mathrm{eff}}^{\mathrm{PCA}}$ and [Fe/H]$^{\mathrm{PCA}}$, are the medians of the distribution. The final uncertainties associated with the atmospheric parameters are the propagation of the residual errors of the calibrations and the standard deviations of the distributions, but [Fe/H]$^{\mathrm{PCA}}$ has an extra term related to the uncertainty of the linear correction term (Equations \ref{eq:error_mc_teff} and \ref{eq:error_mc_feh}).
\begin{equation}
    \sigma \text{T}_{\text{eff}}^{\text{PCA}}= \sqrt{81.26^2 + \sigma_{\text{dist}}^2}
\label{eq:error_mc_teff}
\end{equation}
\begin{equation}
    \sigma \text{[Fe/H]}^{\text{PCA}} = \sqrt{0.1147^2 + \sigma_{\text{dist}}^2 + 0.0762^2}
\label{eq:error_mc_feh}
\end{equation}

This approach to error estimation of the final atmospheric parameters is interesting because it takes into account the position of each spectrum (i.e., the measurements of the EWs of each spectral index) in the parameter space. What is does not take into account is the specific S/N ratio of a given spectrum, since we are considering only a mean value for each spectral index uncertainty. However, we think this simplified approach does not introduce any large uncertainty since the S/N distribution of the spectra of the calibrating stars is very similar to the overall distribution of S/N ratios.

\begin{figure}
\centering
	\includegraphics[width=.75\columnwidth]{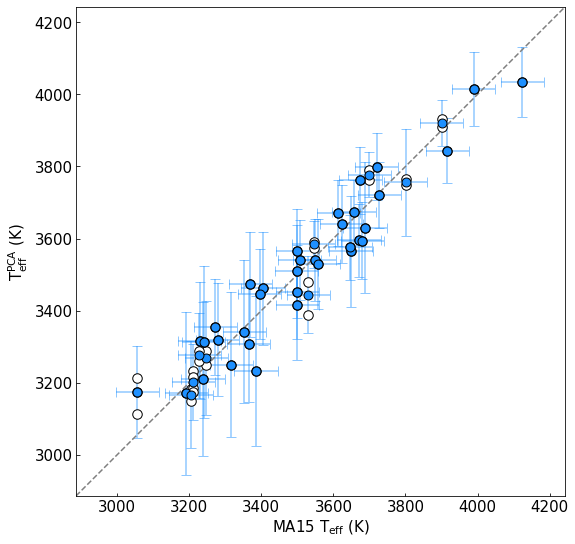}\\
	\includegraphics[width=.75\columnwidth]{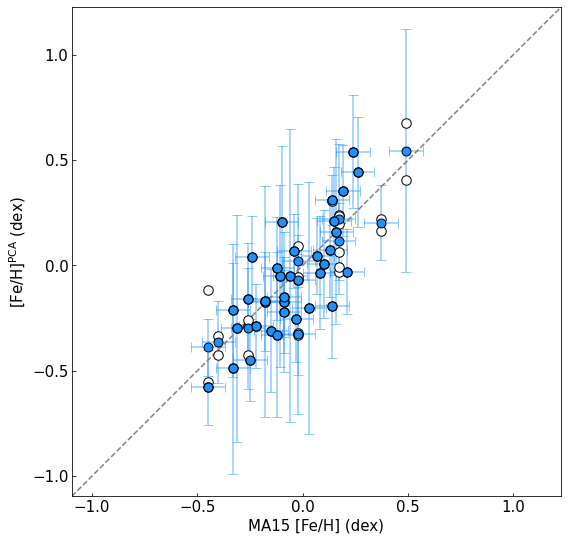}
    \caption{Comparison between the atmospheric parameters from MA15 and the final values from the PCA-based calibrations using MC simulations. The dashed grey lines represent equality. The white circles are the atmospheric parameters found for each spectrum and the blue ones are the final values (weighted mean) representing each star. We have 44 calibration stars for which 10 have more than one spectrum, giving a total of 57 spectra. We found mean differences (related to the blue dots) of 0$\pm$64 K and 0.02$\pm$0.15 dex for T$_{\text{eff}}$ and [Fe/H], respectively.}
    \label{fig:calibration_stars_mc}
\end{figure}

\begin{figure}
	\includegraphics[width=\columnwidth]{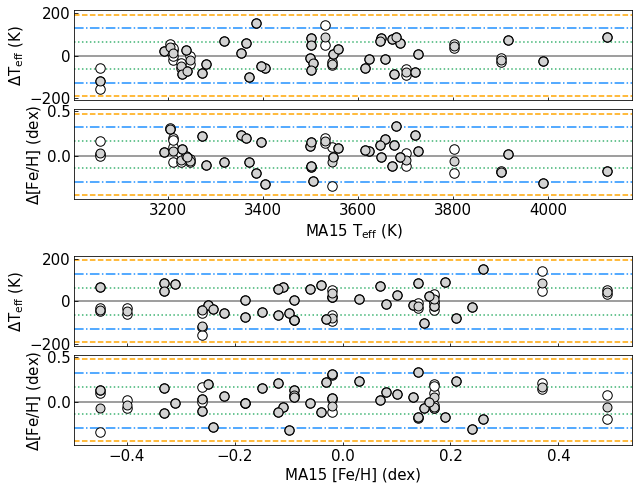}
    \caption{$T_{\mathrm{eff}}^{\mathrm{PCA}}$ and [Fe/H]$^{\mathrm{PCA}}$ residuals versus $T_{\mathrm{eff}}$ and [Fe/H] from MA15 in the upper and bottom panels, respectively. The white circles are the atmospheric parameters found for each spectrum and the grey ones are the final value (weighted mean) representing each star. We have 44 calibration stars for which 10 have more than one spectrum, giving a total of 57 spectra. We found mean differences (grey circles) of 0$\pm$64 K and 0.02$\pm$0.15 dex for $T_{\mathrm{eff}}$ and [Fe/H], respectively. Our residuals reach at most 2$\sigma$ approximately for both atmospheric parameters, the worst cases being the metal-richer and hotter K dwarf stars. The grey solid lines represent the zero, green dotted lines are $\pm$1$\sigma$, blue dash-dotted lines are $\pm$2$\sigma$ and orange dashed lines are $\pm$3$\sigma$ of the distributions.}
    \label{fig:residuals_calibration_mc}
\end{figure}

We applied this procedure to the calibration stars in order to verify the consistency of the atmospheric parameters obtained with the PCA calibrations using MC simulations. As shown in Figure \ref{fig:calibration_stars_mc}, we satisfactorily recovered the atmospheric parameters values from \cite{Mann2015} and found a mean difference of 0$\pm$64 K and 0.02$\pm$0.15 dex and a median difference of -3 K and 0.01 dex for $T_{\mathrm{eff}}$ and [Fe/H], respectively. Greater uncertainties for [Fe/H]$^{\mathrm{PCA}}$ were expected considering that [Fe/H] is difficult to derive due to its low sensitivity to the spectral indices, i.e., it is represented by lower PCs. Furthermore, we stress that, even though we removed all three indices corresponding to the Ca II triplet, chromospheric activity might still be affecting other spectral indices. This is unlikely since the two stronger lines of the Ca II triplet, $\lambda$8498 ($i$60) and $\lambda$8662 ($i$105), are the most opaque spectral features in our coverage (see Figure \ref{fig:example_spectra}), excepting for the very cool types. Chromospheric fill-in in the triplet lines would have to be very strong to appear in other features, and only for the most active stars in our sample does the filling become appreciable. If spectral indices are affected by chromospheric activity, our results will only have a greater dispersion but this should not affect the accuracy -- since the PCA technique minimizes the contributions of parameters that do not have a great impact on the total form of the spectrum. The residuals for both atmospheric parameters are shown in Figure \ref{fig:residuals_calibration_mc} and we found a similar trend with the residuals previously discussed in Sect. \ref{sec:calibrations_by_PCA}. Values for the final atmospheric parameters of the calibration sample are given in Table \ref{tab:literature_calibration}.

\begin{table*}
    \centering
    \caption{Final atmospheric parameters for the calibration stars obtained by the PCA-based MC distributions. Column 1 lists the identifiers used for the stars. Columns 2 and 3 list the coordinates in J2000. Columns 4, 5 and 6 list $T_{\mathrm{eff}}^{\text{MA15}}$ and $T_{\mathrm{eff}}$ and uncertainties from this work, respectively. Columns 7, 8 and 9 list [Fe/H]$^{\text{MA15}}$ and [Fe/H] and uncertainties from this work, respectively. Columns 10 and 11 list the calculated radial velocities and uncertainties. Column 12 lists nominal S/N of the spectra. Column 13 lists spectral types from MA15. $T_{\mathrm{eff}}^{\text{MA15}}$ and [Fe/H]$^{\text{MA15}}$ uncertainties are 60 K and 0.08 dex, respectively, according to other authors.}
    \resizebox{16cm}{!}{
        \begin{tabular}{lcccccccccccc}
        \hline
        \hline
        Name & RA & DEC & $T_{\mathrm{eff}}^{\text{MA15}}$ & $T_{\mathrm{eff}}^{\text{PCA}}$ & $\sigma T_{\mathrm{eff}}^{\text{PCA}}$ & [Fe/H]$^{\text{MA15}}$ & [Fe/H]$^{\text{PCA}}$ & $\sigma$[Fe/H]$^{\text{PCA}}$ & RV & $\sigma$RV & S/N & SpT$^{\text{MA15}}$ \Tstrut\\
        & J2000 & J2000 & (K) & (K) & (K) & (dex) & (dex) & (dex) & (km/s) & (km/s) & & \\
        \hline
        HIP 5643 & 01:12:30.60 & -16:59:56.30 & 3056$\pm$60 & 3174 & 128 & -0.26$\pm$0.08 & -0.3 & 0.29 & 31.18 & 1.04 & 134.70.80 & M4.9 \\ 
        HIP 8768 & 01:52:49.10 & -22:26:05.40 & 3900$\pm$60 & 3921 & 63 & 0.14$\pm$0.08 & 0.31 & 0.12 & 13.23 & 0.96 & 333.251 & M0.2 \\ 
        HIP 12781 & 02:44:15.50 & 25:31:24.10 & 3405$\pm$60 & 3461 & 157 & -0.10$\pm$0.08 & 0.21 & 0.36 & 39.22 & 1.18 & 80 & M3.0 \\ 
        HIP 21556 & 04:37:41.80 & -11:02:19.90 & 3671$\pm$61 & 3594 & 101 & -0.04$\pm$0.08 & 0.07 & 0.18 & -4.84 & 0.87 & 260 & M2.0 \\ 
        HIP 21932 & 04:42:55.70 & 18:57:29.30 & 3680$\pm$60 & 3594 & 107 & 0.14$\pm$0.08 & -0.19 & 0.25 & 17.83 & 0.64 & 130 & M2.2 \\ 
        HIP 22762 & 04:53:49.90 & -17:46:24.30 & 3506$\pm$60 & 3540 & 111 & -0.24$\pm$0.08 & 0.04 & 0.2 & -13.28 & 1.02 & 274 & M2.1 \\ 
        HIP 23512 & 05:03:20.00 & -17:22:24.70 & 3365$\pm$60 & 3307 & 161 & -0.12$\pm$0.08 & -0.33 & 0.39 & 16.86 & 2.15 & 100 & M3.2 \\ 
        HIP 25878 & 05:31:27.30 & -03:40:38.00 & 3801$\pm$60 & 3755 & 147 & 0.49$\pm$0.08 & 0.54 & 0.58 & 11.01 & 0.54 & 220.595 & M1.5 \\ 
        HIP 36208 & 07:27:24.40 & 05:13:32.83   & 3317$\pm$60 & 3250 & 200 & -0.11$\pm$0.08 & -0.05 & 0.43 & 14.08 & 0.73 & 210 & M3.8 \\
        HIP 40501 & 08:16:07.90 & 01:18:09.20 & 3500$\pm$60 & 3566 & 116 & -0.12$\pm$0.08 & -0.01 & 0.24 & 75.43 & 0.93 & 190 & M2.2 \\ 
        HIP 49986 & 10:12:17.60 & -03:44:44.30 & 3623$\pm$60 & 3640 & 107 & 0.13$\pm$0.08 & 0.08 & 0.22 & 20.61 & 0.95 & 180 & M1.9 \\ 
        HIP 51007 & 10:25:10.80 & -10:13:43.20 & 3700$\pm$60 & 3776 & 62 & -0.02$\pm$0.08 & 0.02 & 0.13 & 21.06 & 1.2 & 199.233 & M1.5 \\ 
        HIP 51317 & 10:28:55.50 & 00:50:27.60 & 3548$\pm$60 & 3540 & 113 & -0.18$\pm$0.08 & -0.17 & 0.23 & 20.65 & 0.8 & 170 & M2.2 \\ 
        HIP 53020 & 10:50:52.00 & 06:48:29.20 & 3238$\pm$60 & 3211 & 214 & 0.16$\pm$0.08 & 0.16 & 0.44 & 1.41 & 2.75 & 150 & M3.9 \\ 
        HIP 57548 & 11:47:44.30 & 00:48:16.40 & 3192$\pm$60 & 3170 & 225 & -0.02$\pm$0.08 & -0.07 & 0.45 & -33.02 & 1.12 & 190 & M4.3 \\ 
        GJ 3707 & 12:10:05.60 & -15:04:16.90 & 3385$\pm$60 & 3232 & 207 & 0.26$\pm$0.08 & 0.44 & 0.26 & 78.64 & 0.77 & 218 & M3.8 \\ 
        HIP 62687 & 12:50:43.50 & -00:46:05.20 & 3989$\pm$60 & 4015 & 103 & 0.24$\pm$0.08 & 0.54 & 0.27 & 0.08 & 1.53 & 305 & K7.9 \\ 
        GJ 512a & 13:28:21.00 & -02:21:37.10 & 3498$\pm$60 & 3509 & 129 & 0.08$\pm$0.08 & -0.03 & 0.27 & -38.74 & 1.92 & 283 & M3.1 \\ 
        HIP 65859 & 13:29:59.70 & 10:22:37.70 & 3727$\pm$61 & 3721 & 91 & -0.09$\pm$0.08 & -0.15 & 0.19 & 30.45 & 1.24 & 160 & M1.1 \\ 
        HIP 67155 & 13:45:43.70 & 14:53:29.40 & 3649$\pm$60 & 3565 & 154 & -0.31$\pm$0.08 & -0.3 & 0.54 & 39.34 & 1.29 & 190 & M1.4 \\ 
        HIP 71253 & 14:34:16.80 & -12:31:10.40 & 3211$\pm$60 & 3201 & 105 & 0.17$\pm$0.08 & 0.11 & 0.18 & 1.55 & 0.71 & 70.233.130.180 & M4.0 \\ 
        HIP 74995 & 15:19:26.80 & -07:43:20.10 & 3395$\pm$60 & 3446 & 125 & -0.15$\pm$0.08 & -0.31 & 0.29 & 5.06 & 1.05 & 180 & M3.2 \\ 
        HIP 80824 & 16:30:18.00 & -12:39:45.30 & 3272$\pm$60 & 3354 & 133 & -0.03$\pm$0.08 & -0.25 & 0.2 & -21.91 & 0.87 & 437 & M3.6 \\ 
        HIP 82809 & 16:55:25.20 & -08:19:21.30 & 3279$\pm$60 & 3318 & 157 & -0.26$\pm$0.08 & -0.16 & 0.26 & 11.11 & 1.56 & 232 & M3.2 \\ 
        GJ 1207 & 16:57:05.70 & -04:20:56.30 & 3229$\pm$60 & 3316 & 163 & -0.09$\pm$0.08 & -0.17 & 0.33 & -3.16 & 1.29 & 232 & M4.1 \\ 
        HIP 85295 & 17:25:45.20 & 02:06:41.10 & 4124$\pm$60 & 4034 & 96 & 0.19$\pm$0.08 & 0.35 & 0.21 & -26.64 & 2.77 & 180 & K7.4 \\ 
        HIP 85665 & 17:30:22.70 & 05:32:54.60 & 3675$\pm$60 & 3761 & 92 & -0.09$\pm$0.08 & -0.22 & 0.2 & -15.87 & 1.88 & 150 & M0.5 \\ 
        HIP 86287 & 17:37:53.30 & 18:35:30.10 & 3657$\pm$60 & 3672 & 91 & -0.25$\pm$0.08 & -0.45 & 0.19 & -12.53 & 1.62 & 140 & M1.2 \\ 
        HIP 87937 & 17:57:48.40 & 04:41:36.10 & 3228$\pm$60 & 3275 & 118 & -0.40$\pm$0.08 & -0.36 & 0.2 & -111.21 & 0.47 & 520.18 & M4.2 \\ 
        HIP 88574 & 18:05:07.50 & -03:01:52.70 & 3614$\pm$60 & 3670 & 92 & -0.22$\pm$0.08 & -0.29 & 0.2 & 32.08 & 1.1 & 160 & M1.3 \\ 
        HIP 92403 & 18:49:49.30 & -23:50:10.40 & 3240$\pm$60 & 3313 & 212 & -0.18$\pm$0.08 & -0.17 & 0.55 & -11.94 & 1.07 & 180 & M4.1 \\ 
        HIP 93873 & 19:07:05.50 & 20:53:16.90 & 3500$\pm$60 & 3452 & 130 & -0.33$\pm$0.08 & -0.21 & 0.32 & 32.55 & 1.42 & 218 & M2.1 \\ 
        HIP 93899 & 19:07:13.20 & 20:52:37.20 & 3494$\pm$62 & 3415 & 152 & -0.35$\pm$0.08 & -0.49 & 0.5 & 34.44 & 2.42 & 90 & M2.1 \\ 
        HIP 94761 & 19:16:55.20 & 05:10:08.00 & 3558$\pm$60 & 3528 & 125 & 0.10$\pm$0.08 & 0.01 & 0.26 & 34.67 & 0.73 & 200 & M2.6 \\ 
        HIP 103039 & 20:52:33.00 & -16:58:29.00 & 3205$\pm$60 & 3166 & 149 & -0.02$\pm$0.08 & -0.33 & 0.38 & 15.87 & 1.17 & 130.8 & M4.0 \\ 
        HIP 104432 & 21:09:17.40 & -13:18:09.00 & 3545$\pm$60 & 3584 & 65 & -0.45$\pm$0.08 & -0.39 & 0.14 & -61.6 & 1.29 & 223.139 & M1.4 \\ 
        HIP 109388 & 22:09:40.30 & -04:38:26.60 & 3530$\pm$60 & 3443 & 106 & 0.37$\pm$0.08 & 0.2 & 0.18 & -14.22 & 0.6 & 294.17 & M3.1 \\ 
        HIP 111571 & 22:36:09.60 & -00:50:29.70 & 3916$\pm$61 & 3843 & 90 & 0.07$\pm$0.08 & 0.04 & 0.18 & 8.07 & 1.89 & 250 & M0.6 \\ 
        HIP 113020 & 22:53:16.70 & -14:15:49.30 & 3247$\pm$60 & 3268 & 156 & 0.17$\pm$0.08 & 0.22 & 0.36 & 0.02 & 0.61 & 170.15 & M3.7 \\ 
        HIP 113296 & 22:56:34.80 & 16:33:12.30 & 3720$\pm$60 & 3799 & 93 & 0.21$\pm$0.08 & -0.03 & 0.2 & -28.81 & 1.17 & 110 & M1.5 \\ 
        HIP 114046 & 23:05:52.00 & -35:51:11.00 & 3688$\pm$86 & 3630 & 182 & -0.06$\pm$0.08 & -0.05 & 0.7 & 24.04 & 1.48 & 200 & M1.1 \\ 
        GJ 896a & 23:31:52.10 & 19:56:14.10 & 3353$\pm$60 & 3341 & 198 & 0.03$\pm$0.08 & -0.2 & 0.6 & 0.89 & 2.39 & 102 & M3.8 \\ 
        HIP 117473 & 23:49:12.50 & 02:24:04.40 & 3646$\pm$60 & 3576 & 90 & -0.45$\pm$0.08 & -0.58 & 0.18 & -71.04 & 1.21 & 150 & M1.4 \\ 

        \hline
        \end{tabular}
    }
\label{tab:literature_calibration}
\end{table*}

\begin{figure}
	\includegraphics[width=\columnwidth]{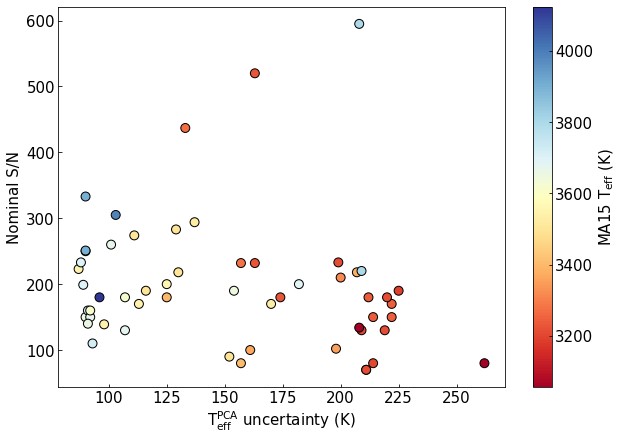}\\
	\includegraphics[width=\linewidth]{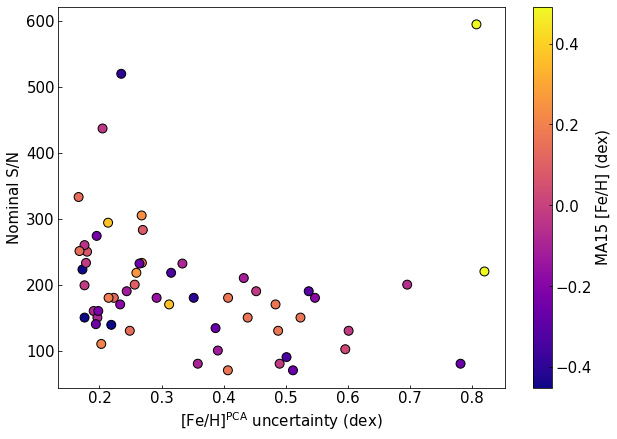}
    \caption{Correlation between the atmospheric parameters of MA15 and our T$_{\text{eff}}^{\text{PCA}}$ and [Fe/H]$^{\text{PCA}}$ values, with their respective uncertainties and nominal S/N on top and bottom panels, respectively, for 57 spectra of 44 calibration stars. The hot outlier on the top panel with uncertainty around 210 K is the spectrum of HIP 25878 (M1.5Ve) and this difference might be due to its extremely high S/N of 600 maybe introducing difficulties in applying the same index system based on lower S/N spectra.}
    \label{fig:sn_uncertainties}
\end{figure}

We explored the correlation between \cite{Mann2015} atmospheric parameters, the uncertainties of the final atmospheric parameters delivered by our PCA calibration and the nominal S/N of the spectra and found a strong trend for T$_{\text{eff}}^{\text{PCA}}$ residuals. Figure \ref{fig:sn_uncertainties} shows that cooler stars have lower S/N due to the exposure times of these objects (see Table \ref{tab:exptime}), but even for cooler stars there is a decrease of the average uncertainties when we achieve higher S/N. On the other hand, [Fe/H] values are better distributed between the PCA-based uncertainties and S/N, indicating that there are no systematic errors associated with the [Fe/H]values themselves, i.e., higher metallicities do not necessarily have higher uncertainties.

We applied the same procedure to the 134 stars which are not part of the calibration sample (hereafter called ``study sample''). For this case, all values outside the parameter space of the calibration sample are extrapolating the calibrations and can not provide trustful values, yet slight extrapolations outside the parameter space of the calibrating stars should not introduce any large errors. 


\subsection{Comparison with literature values}

\begin{figure*}
    \centering
    \resizebox{16cm}{!}{\includegraphics{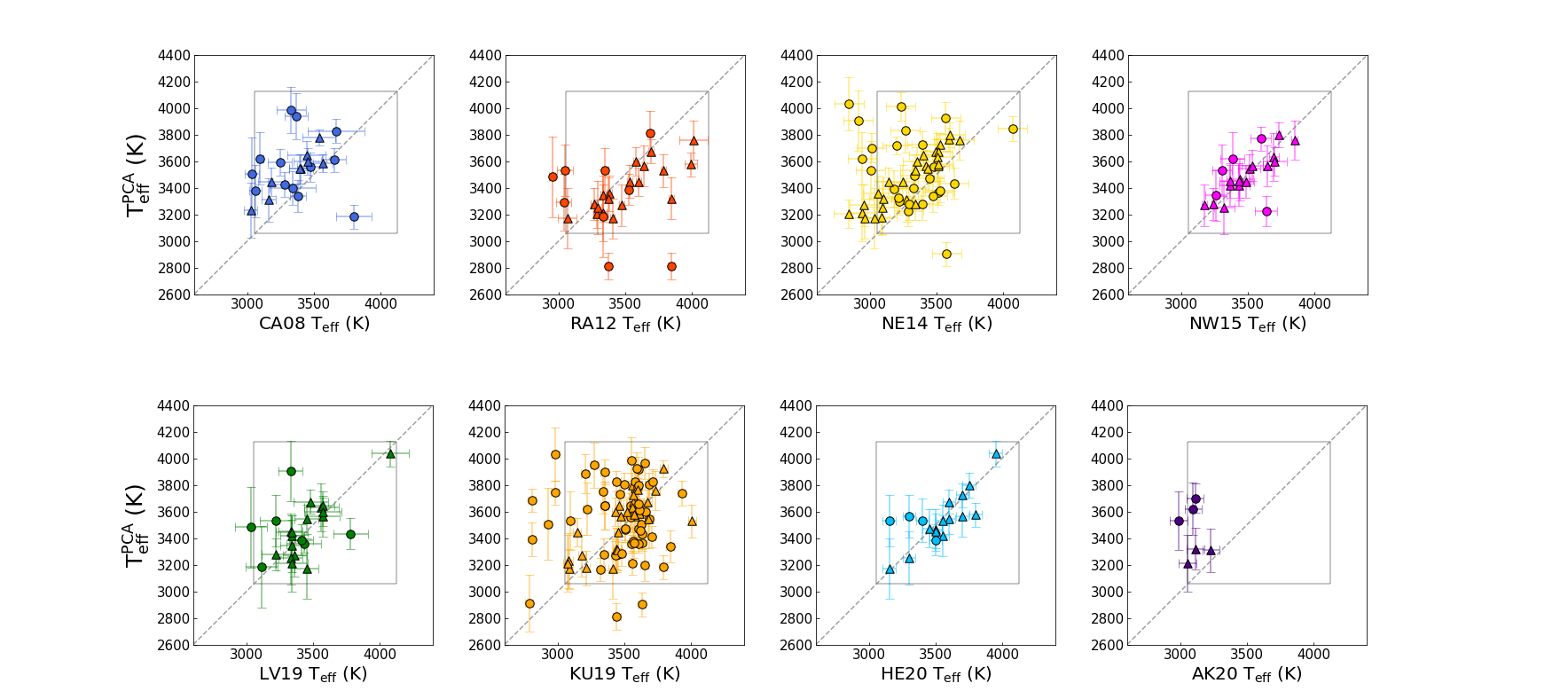}}
    \caption{Comparison between our T$_\text{eff}\text{PCA}$ for stars in commom with other studies. The grey dashed lines represent equality and the boxes represent the limits of strict validity of our calibrations. The triangles represent the calibration stars and the circles represent sample stars -- see text for details.}
    \label{fig:teff_literature}
\end{figure*}

\begin{figure*}
    \centering
    \resizebox{16cm}{!}{\includegraphics{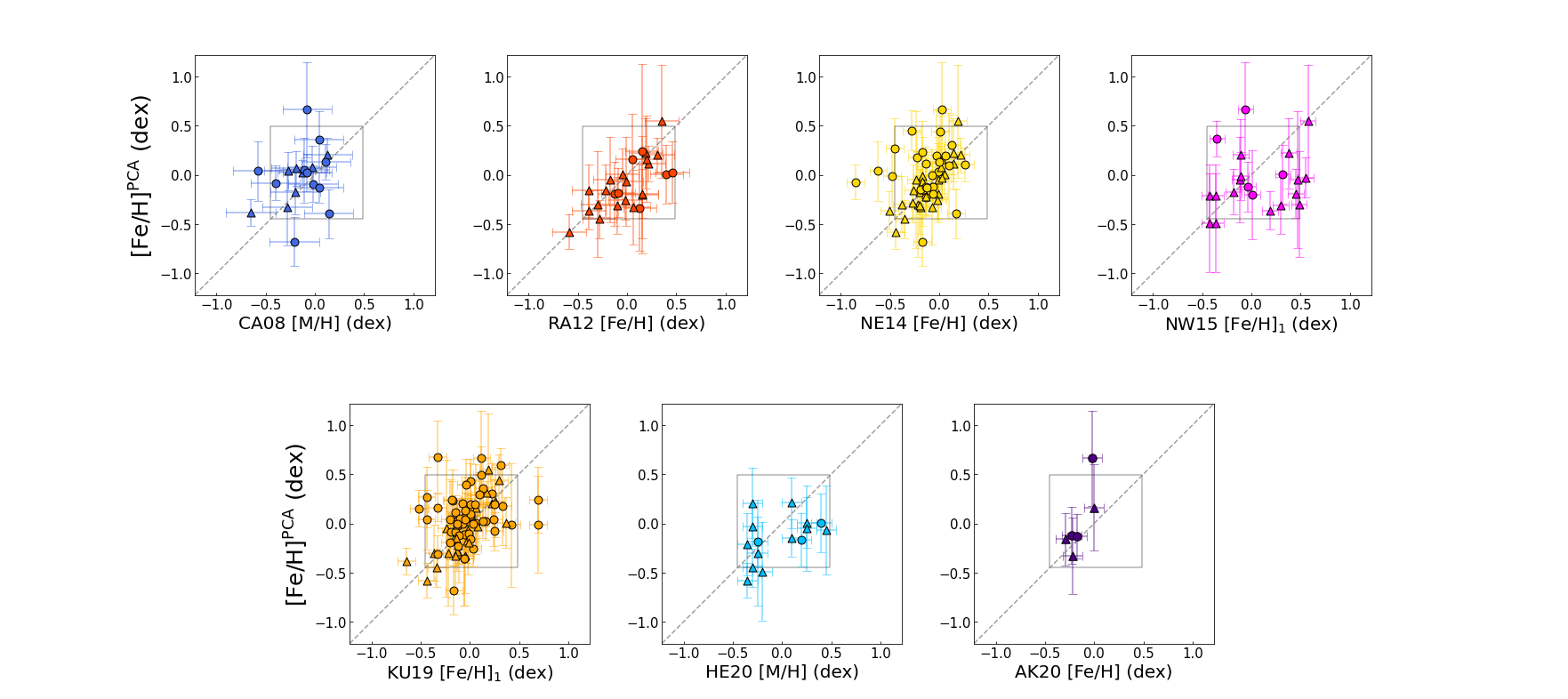}}
    \caption{Comparison between our [Fe/H]$^\text{PCA}$ for stars in commom with other studies. The grey dashed lines represent equality and the boxes represent the limits of strict validity of our calibrations. The triangles represent the calibration stars and the circles represent sample stars -- see text for details.}
    \label{fig:feh_literature}
\end{figure*}

We compare our final atmospheric parameters with a variety of published studies employing photometic and spectroscopic determinations for a wider appraisal of the accuracy and precision of our approach. \cite{Casagrande2008} (hereafter, CA08) exploits the flux ratio in different optical-infrared bands as a sensitive indicator of atmospheric parameters\footnote{Quoted from \citealt{Casagrande2008}: ``The model atmospheres are given for the total heavy elements, but for low values of alpha–enhancement, the difference between the two is negligible, particularly since metallicity measurements in M dwarfs are still uncertain.''.}; \cite{Rojas-Ayala2012} (hereafter, RA12) estimates [Fe/H] from Na I, Ca I, and H2O-K2 measurements and used the H2O-K2 index as an indicator of T$_{\text{eff}}$; \cite{Neves2014} (hereafter, NE14) estimations were based on the measurement of pseudo equivalent widths of features in the 530-690 nm range of HARPS spectra; \cite{Newton2015} (hereafter, NW15) T$_{\text{eff}}$ are based on EWs of H-band spectral features and [Fe/H] are derived following \citealt{Mann2013a} calibrations ([Fe/H]$_{1}$); \cite{Lopez-Valdivia2019} (hereafter, LV19) uses high-resolution line-depths measurements in the H-band from Grating Infrared Spectrometer (IGRINS) spectra to estimate T$_{\text{eff}}$; \cite{KU19} (hereafter, KU19) determines atmospheric parameters by fitting the observed HARPS spectra with a grid of BT-Settl stellar atmosphere models; \cite{HE20} (hereafter, HE20) uses the BT-Settl model atmospheres to derive atmospheric parameters of low- to medium-resolution spectra; \cite{AK20} (hereafter, AK20) uses a machine learning tool (ODUSSEAS) to derive atmospheric parameters based on the measurement of the pseudo equivalent widths for more than 4000 stellar absorption lines. The comparisons are shown in Figures \ref{fig:teff_literature} and \ref{fig:feh_literature} and the following discussion is limited to stars lying within the parameter space of our calibrations. 


Regarding T$_{\text{eff}}$, we notice that our values have a systematic overestimation of 186$\pm$255 K, 349$\pm$206 K and 219$\pm$271 K compared to CA08, NE14 and AK20, respectively. NE14 and AK20 estimations were calibrated following CA08 values (explaining their mutual consistency) which uses black-body approximations for M dwarfs in the infrared. Depression of the local pseudo continuum by molecular line blanketing may have led to an underestimation of the flux and hence an underestimation of T$_{\text{eff}}$. We found an excellent agreement of 11$\pm$133 K and 47$\pm$199 K for T$_{\text{eff}}$ between our values and NW15 and LV19, respectively, both based on H-band spectral features, and -40$\pm$227 K with RA12 and 12$\pm$140 K with HE20. Lastly, we found a difference of 84$\pm$309 K with KU19, showing a significant zero-point discrepancy but contained by a large dispersion.
This exceptional agreement with NW15 values were expected since they based their work on interferometric measurements in common with MA15. We recall that the mean and median T$_{\text{eff}}^{\text{PCA}}$ uncertainties are 132 K and 105 K, respectively, for the study sample. From Figure \ref{fig:teff_literature} it is noteworthy that, even in the middle of the parameter range, T$_{\text{eff}}$ offsets up to many hundreds of Kelvin are seen between published values, for example, between the barycenters of the CA08 and RA12 T$_{\text{eff}}$ distributions. Our values tend to lie midrange in the distribution of published T$_{\text{eff}}$ values.


Regarding [Fe/H], we achieve good agreement with CA08, RA12, NE14, KU19 and HE20 with mean differences of 0.11$\pm$0.24 dex, -0.11$\pm$0.19 dex, 0.09$\pm$0.26 dex, 0.07$\pm$0.33 dex and -0.02$\pm$0.25 dex, respectively, meaning that possibly not all scatter in these comparisons is due to our calibration (see also Table \ref{tab:literature}). Although NW15 [Fe/H]$_{1}$ is based on the same calibration used by \citealt{Mann2015} applicable to stars ranging -1.04$<$[Fe/H]$<$+0.56 and spectral types from K7 to M5, we found one of the worst correlation of the comparisons with a mean difference of -0.16$\pm$0.38 dex between theirs and our calibrated values. We found a weak correlation of 0.17$\pm$0.25 with AK20, but we note that there are only 6 stars in common with our sample. We recall that the mean and median [Fe/H]$^{\text{PCA}}$ uncertainties are 0.28 dex and 0.23 dex, respectively, for the study sample. From Figure \ref{fig:feh_literature} we see that agreement for published [Fe/H] is much better than for T$_{\text{eff}}$, and again our determinations lie midrange. Values for the final atmospheric parameters of 134 stars of the study sample are given in Table \ref{tab:literature_study_sample}.


\subsection{Appraisal of discrepancies between different sources}
We next perform a more detailed overall comparison between the atmospheric parameters from different works: the results are shown in Table \ref{tab:literature}. This comparison considers analyses employing very different methods and heterogeneous sources of data. Intercomparisons comprising less than 10 stars are not statistically significant and will not be pursued. We begin by comparing T$_{\text{eff}}$ values. Concerning our own determinations, it is apparent that intercomparisons involving our values tend to have large dispersions, though not exclusively. Important offsets are seen between our work and CA08 and NE14, even though contained within the large dispersions: our values are hotter in both cases. Large offsets are also seen when comparing CA08 to all authors but NE14 and AK20, as expected, since the latters' T$_{\text{eff}}$ determinations were made consistent to the former's. Particularly significant are the discrepancies between the values of CA08-NE14 and those of RA12, NW15 and HE20, as the offsets clearly surpass the observed dispersions by large margins. The determinations of MA15, RA12, NW15 and LV19 are mutually consistent. Besides those that involve the values of this work, very large dispersions are observed in the comparisons RA12/NE14, RA12/LV19, NE14/LV19, CA08/KU19, RA12/KU19, LV19/KU, CA08/HE20, MA15/HE20, NW15/HE20, LV19/HE20 and KU19/HE20 though not always associated with large offsets.

Intercomparisons involving [Fe/H] determinations fare much better. Again, the largest dispersions are found when comparing our own values with those of other works, but the offsets are well contained by the dispersions. The largest statistically significant offsets in [Fe/H] are seen between CA08/MA15, CA08/KU19, CA08/HE20 and NW15/HE20, the first one largely surpassing the dispersion. All other comparisons show good accord, and the dispersions are generally low, with some exceptions. The fact that some [Fe/H] comparisons between sources with more precise determinations than our own show dispersions in the $>$0.10 dex range seems to justify our previous assertion that not all scatter seen in Figure \ref{fig:feh_literature} is due to our method.

Our atmospheric parameters are overall quite accurate but not very precise. Nonetheless, they are competitive with estimations that make use of high-resolution spectroscopy. We find a median difference of 75$\pm$273 K and 0.02$\pm$0.31 dex for T$_{\text{eff}}$ and [Fe/H], respectively.
Lastly, considering the significant number of stars in common between previously published studies, and the fact that most stars in common between this work and other authors' belong to the calibration sample itself (i.e., stars in common with \citealt{Mann2015}), it is noteworthy that here we present determinations of atmospheric parameters for a large number of M dwarfs for which previously such data were lacking.

\begin{table*}
    \centering
    \caption{Mean differences of effective temperature and metallicity between our work and other authors. The right diagonal contains the number of stars in common between authors and the left diagonal contains the mean differences and standard deviations for the atmospheric parameters in K and dex. Differences are calculated in the following way: values of the parameters from the authors listed in the top row minus those values from the authors listed in the leftmost column. Comparisons containing 10 stars or less in common between authors should be considered with caution. See text for more details.}
    \begin{threeparttable}
    \resizebox{16cm}{!}{
        \begin{tabular}{c|ccccccccccc}
        \hline
        \hline
        \diagbox[height=0.7cm, width=2.2cm]{Ref.}{Ref.} & Costa-Almeida & CA08 & RA12 & NE14 & MA15 & NW15 & LV19 & KU19 & HE20 & AK20 \\ 
        \hline
        \multirow{2}{*}{Costa-Almeida} &  & \multirow{2}{*}{22} & \multirow{2}{*}{27} & \multirow{2}{*}{52} & \multirow{2}{*}{44\tnote{a}} & \multirow{2}{*}{21} & \multirow{2}{*}{23} & \multirow{2}{*}{81} & \multirow{2}{*}{18} & \multirow{2}{*}{6} \\
        & & & & & & & & & & \\ \hline
        \multirow{2}{*}{CA08} & 186$\pm$255 &  & \multirow{2}{*}{3} & \multirow{2}{*}{14} & \multirow{2}{*}{19} & \multirow{2}{*}{3} & \multirow{2}{*}{10} & \multirow{2}{*}{81} & \multirow{2}{*}{10} & \multirow{2}{*}{3}\\
        & 0.11$\pm$0.24 & & & & & & & & & \\ \hline
        \multirow{2}{*}{RA12} & -40$\pm$227 & -354$\pm$50 & & \multirow{2}{*}{26} & \multirow{2}{*}{62} & \multirow{2}{*}{26} & \multirow{2}{*}{50} & \multirow{2}{*}{19} & \multirow{2}{*}{0} & \multirow{2}{*}{7}\\
        & -0.11$\pm$0.19 & -0.20$\pm$0.02 & & & & & & & & \\ \hline
        \multirow{2}{*}{NE14} & 210$\pm$278 & 12$\pm$70 & 231$\pm$209 & & \multirow{2}{*}{38} & \multirow{2}{*}{15} & \multirow{2}{*}{20} & \multirow{2}{*}{61} & \multirow{2}{*}{0} & \multirow{2}{*}{30} \\
        & 0.09$\pm$0.26 & -0.08$\pm$0.06 & 0.05$\pm$0.11 & & & & & & & \\ \hline
        \multirow{2}{*}{MA15} & 0$\pm$66\tnote{a} & -143$\pm$102 & 36$\pm$124 & -174$\pm$119 & & \multirow{2}{*}{36} & \multirow{2}{*}{53} & \multirow{2}{*}{55} & \multirow{2}{*}{56} & \multirow{2}{*}{9} \\
        & 0.02$\pm$0.15\tnote{a} & -0.15$\pm$0.06 & -0.04$\pm$0.09 & -0.07$\pm$0.09 & & & & & \\ \hline
        \multirow{2}{*}{NW15} & 11$\pm$133 & -274$\pm$130 & -27$\pm$102 & -161$\pm$169 & 16$\pm$76 & & \multirow{2}{*}{32} & \multirow{2}{*}{16} & \multirow{2}{*}{87} & \multirow{2}{*}{8} \\
        & -0.16$\pm$0.38 & -0.02$\pm$0.00 & -0.01$\pm$0.09 & -0.05$\pm$0.14 & 0.02$\pm$0.08 & & & & & \\ \hline
        \multirow{2}{*}{LV19} & \multirow{2}{*}{47$\pm$199} & \multirow{2}{*}{-142$\pm$140} & \multirow{2}{*}{-19$\pm$244} & \multirow{2}{*}{-233$\pm$331} & \multirow{2}{*}{11$\pm$118} & \multirow{2}{*}{9$\pm$111} & & \multirow{2}{*}{27} & \multirow{2}{*}{49} & \multirow{2}{*}{3} \\
        & & & & & & & & & & \\ \hline
        \multirow{2}{*}{KU19} & 84$\pm$309 & -74$\pm$225 & 92$\pm$230 & -142$\pm$150 & 42$\pm$164 & -28$\pm$146 & \multirow{2}{*}{21$\pm$288} & & \multirow{2}{*}{32} & \multirow{2}{*}{20} \\
        & 0.07$\pm$0.33 & -0.21$\pm$0.19 & 0.00$\pm$0.12 & -0.06$\pm$0.07 & 0.00$\pm$0.12 & -0.13$\pm$0.19 & & & & \\ \hline
        \multirow{2}{*}{HE20} & 12$\pm$140 & 286$\pm$263 & \multirow{2}{*}{-} & \multirow{2}{*}{-} & 97$\pm$332 & -104$\pm$227 & \multirow{2}{*}{31$\pm$346} & 224$\pm$276 & & \multirow{2}{*}{1} \\
        & -0.02$\pm$0.25 & -0.23$\pm$0.32 & & & 0.05$\pm$0.56 & 0.20$\pm$0.65 & & 0.11$\pm$0.61 & & \\ \hline
        \multirow{2}{*}{AK20} & 349$\pm$206 & -19$\pm$36 & 248$\pm$243 & -12$\pm$101 & 168$\pm$51 & 132$\pm$101 & \multirow{2}{*}{235$\pm$149} & 190$\pm$101 & 226$\pm$0 & \\
        & 0.17$\pm$0.25 & -0.08$\pm$0.02 & 0.12$\pm$0.11 & 0.04$\pm$0.06 & 0.11$\pm$0.06 & 0.10$\pm$0.14 & & 0.08$\pm$0.05 & 0.61$\pm$0 & \\
        
        \hline
        \end{tabular}
    }
\begin{tablenotes}
\footnotesize
    \item[a] Calibration sample
\end{tablenotes}
\end{threeparttable}
\label{tab:literature}
\end{table*}

\subsection{M Dwarf Metallicity Distribution Function}
Measuring the [Fe/H] distribution of stars in the Milky Way disk is a fundamental tool of the study of its chemo-dynamical evolution. Cool dwarf stars from spectral types FGKM preserve in their atmospheres, with high fidelity, the chemical composition of their natal interstellar clouds. Moreover, GK dwarfs have lifetimes comparable to the age of the Galaxy, while lifetimes of M dwarfs vastly surpass the Galactic age: these stars thus provide an uninterrupted fossil record of chemical evolution. These chemical footprints are crucial constraints to the star formation history of the Galactic disk as well as its variation in time and space. Such data track the evolution of relevant physical processes such as kinematics, the structure and speed of enrichment, the mass function, the frequency and yields of core collapse and SNIa supernovae, the infall and outflow of gas, and the merging and accretion history. Spectroscopically derived metallicity distribution functions (MDF hereafter) of [Fe/H] abundances have been extensively studied for the more massive FGK stars but data on the [Fe/H] distribution for M dwarfs is much more scarce owing to the difficulty of volume-sampling these objects to appropriate depths. We thought it worthwhile to compare the [Fe/H] distribution resulting from our M dwarf data from the immediate solar neighbourhood to the existing record for FGK dwarfs. We need to keep in mind that even in the immediate neighborhood incompleteness for magnitude-limited sampling of M dwarfs remains very high, particularly for the later types. \cite{Winters2015} find that the identification of M dwarf systems already loses completeness at distances not much farther than 5 pc.

The local disk MDF has been examined both by photometric determinations of [Fe/H] (\citealt{Helio1998}, \citealt{Nordstrom2004}, \citealt{Casagrande2011}) and by spectroscopic [Fe/H] estimations based on spectra with a wide range of spectral resolutions (\citealt{LH2005}, \citeyear{LH2006}, \citeyear{LH2007}; \citealt{Katz2011}; \citealt{Siebert2011} \citealt{Schelesinger2012}). From the many excellent resources available in the literature we chose to compare the [Fe/H] distribution of our M dwarf sample to the statistically large RAVE survey (\citealt{Boeche2013}) by means of the [Fe/H] determinations of their 4th Data Release (\citealt{Kordopatis2013}). The reasons for this choice are manifold. The RAVE survey sampled over 480,000 stars in the Milky Way at a spectral resolution of R $\sim$ 7,500, for which distances were estimated and individual chemical abundances for six elements determined, of which only Fe concerns us here. RAVE is a magnitude-limited survey with constant integration time, and more distant stars have on average lower S/N spectra, not too dissimilar from our own observing strategy. The analysis of \cite{Boeche2013} limits the investigation to dwarf stars (log g $>$ 3.8 dex) with good quality RAVE spectra (i.e., good chemical abundances) and reliable distances, selecting only spectra with S/N $>$ 40 in the 5250 K $<$ T$_\text{eff}$ $<$ 7000 K range (i.e., early F to early K stars) with a resulting average uncertainty for [Fe/H] of $\sim$ 0.15 dex. Their selection yields 19,962 stars, most within 300 pc of the Galactic plane and in the galactocentric radius interval 7.6 kpc $<$ R$_\text{g}$ < 8.3 kpc (they assume R$_\odot$ = 8.0 and R$_\text{g}$ stands for guiding radius). Therefore, the RAVE survey concentrates on dwarf stars; its data possesses spectral resolution very similar to ours; its abundance errors are also comparable; and most of the RAVE sample is inside the scale height of the thin disk ($~$0.3 kpc, as shown by the Figure 1 of \citealt{Boeche2013}), which is certainly the case for our sample of very nearby stars (see Sect. \ref{sec:dynamics}).

\begin{figure}
    \centering
    \includegraphics[width=\columnwidth]{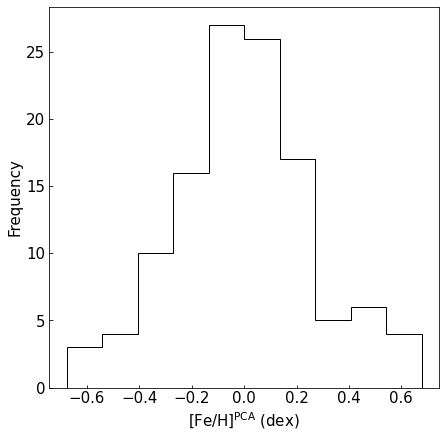}
    \caption{The raw [Fe/H] distribution of all [Fe/H]$_\text{PCA}$ determinations derived in this study, out to 25 pc, binned to 0.2 dex intervals.}
    \label{fig:histogram_feh}
\end{figure}

The [Fe/H] distribution of the full sample of RAVE stars from \cite{Boeche2013} is shown in their Figure 2, and it has a well-defined peak at [Fe/H]  $\sim$ -0.2 dex. Particularly, their Figure 3 shows the distribution for those stars with 0.0 kpc $<\left|\text{Z}_{\text{max}}\right| <$ 0.4 kpc, where $\left|\text{Z}_{\rm \text{max}}\right|$ stands for maximum excursion above the Galactic plane, and this distribution, from a sample fully dominated by thin disk stars, peaks at [Fe/H] of -0.15 dex. Our own [Fe/H] distribution, sampled to 0.2 dex to reflect external [Fe/H] errors, and employing the atmospheric parameters derived from the PCA calibration even for the calibrating stars themselves, for the sake of consistency, is shown in Figure \ref{fig:histogram_feh}. It peaks at [Fe/H] $\sim$ -0.10 dex. Our M dwarf MDF shows, thus, good agreement with that of FGK stars of the local thin disk, within the uncertainties, and we conclude that the MDFs of FGKM stars in the local thin disk are consistent between these spectral types.

\section{Dynamic configuration in the Galactic structures}
\label{sec:dynamics}

\begin{figure*}
    \centering
    \includegraphics[width=0.3\linewidth]{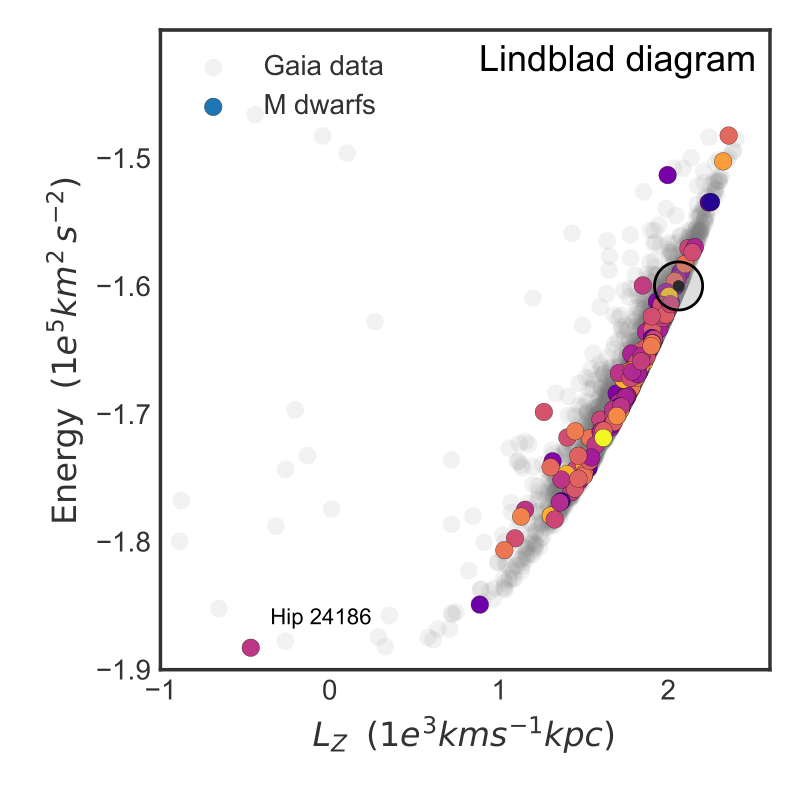}
    \includegraphics[width=0.3\linewidth]{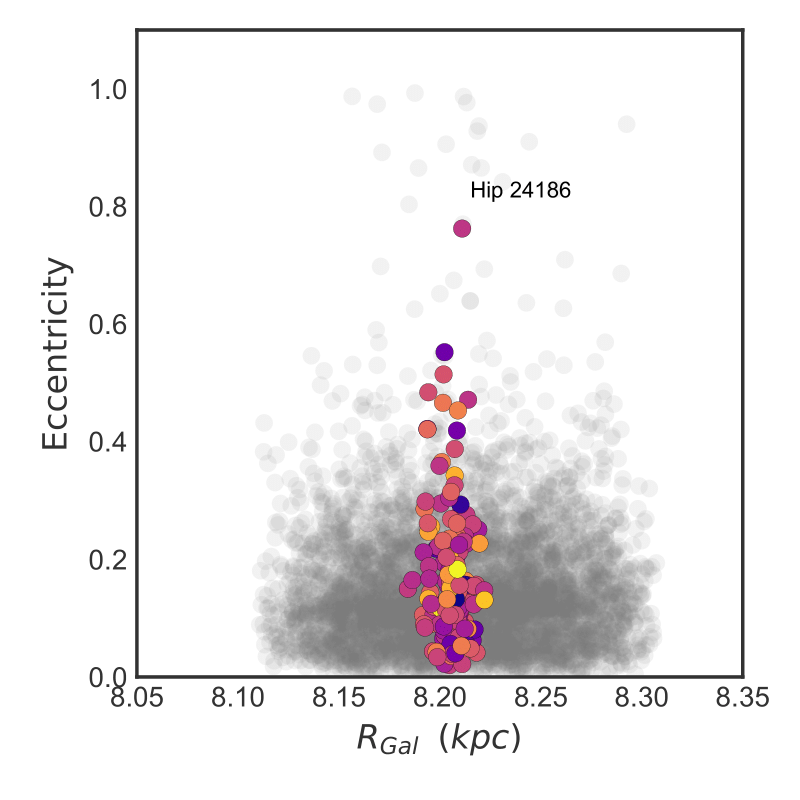}
    \includegraphics[width=0.358\linewidth]{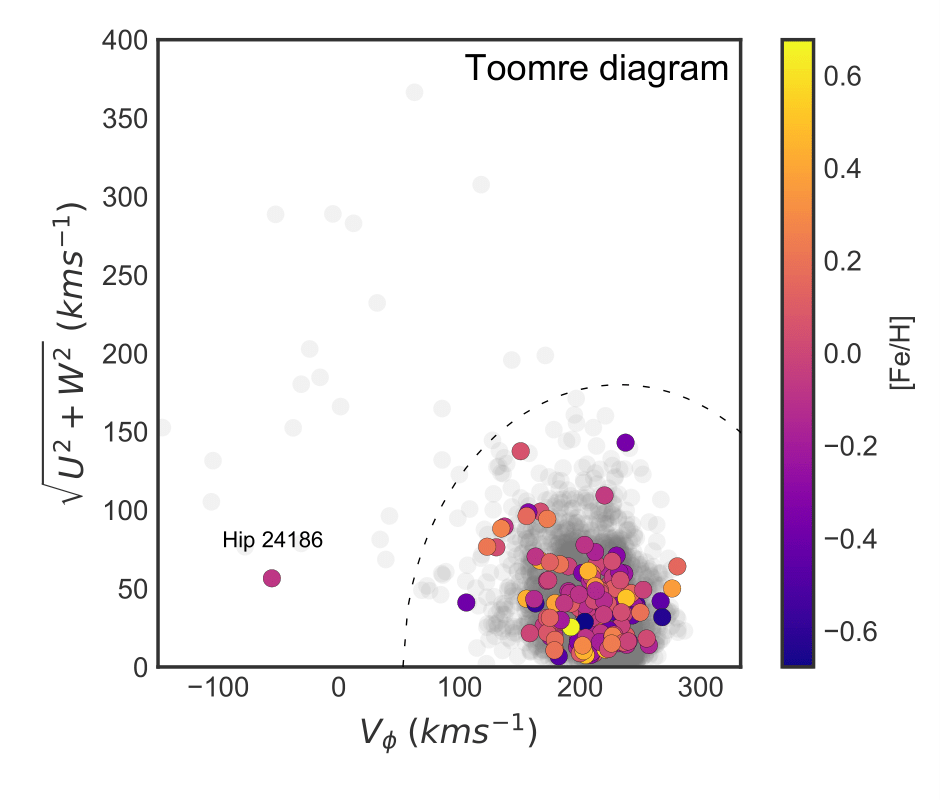}
    \caption{M dwarfs and Gaia random stars within 10~pc in the Limdblad diagram (left panel), in the eccentricity vs. $R_{\mathrm{Gal}}$ diagram (mid panel), and in the Toomre diagram (right panel). 
    Our red dwarf data are color coded according to the bar in the right panel, whereas the Gaia random stars are shown in gray.
    In the left panel, the position of the Sun is indicated with the $\odot$ symbol.
    In the right panel, the dashed line corresponds to $V_{\phi\,\mathrm{total}} = 180\ \text{kms}^{-1}$ with respect to the local standard of rest.}
    \label{fig:kinematics}
\end{figure*}

Figure~\ref{fig:kinematics} shows the distribution of the M dwarfs in diagrams frequently used to determine the membership of the stars in the major Galactic substructures. 
These are the Lindblad and Toomre diagrams, and the eccentricity vs. Galactic radius ($R_{\mathrm{Gal}}$). 
The Gaia nearby (distances within 100 pc) random data, represented by the gray dots, was plotted as background to facilitate the visualisation of the M dwarfs with respect to a representative reference in the solar surrounding.
As expected, most of the stars remain within the areas corresponding to the thin and thick discs \citep[e.g.][among many others]{helmi2000MNRAS.319..657H, gallart2019, massari2019}. 
In the Toomre diagram this area is enclosed by the dashed line, the radius of which was set to $V_{\phi\,\mathrm{total}} = 180\ \text{kms}^{-1}$.
The $V_{\phi\,\mathrm{total}}$ quantity is frequently defined as the velocity module that separates prograde and retrograde movements; 
for example, \citet{bonaca2017ApJ...845..101B} adopts $220\ \text{kms}^{-1}$.
However, it is important to keep in mind that a fixed figure is just a representative value to separate the halo and the discs, the velocity distributions of which may naturally overlap.

HIP 24186 is the only star to stand clearly outside of the areas covered by the disc stars in every diagram. This object is the well-known high-velocity Kapteyn's star. According to our modelling it has a low orbital energy, a slightly retrograde movement $V\approx-50\ \text{kms}^{-1}$, and a quite eccentric orbit $\sim0.75$. It is a moderately metal-rich star for such kinematics, in the range [Fe/H] $\sim$ -0.60 dex to -0.90 dex (e.g., \citealt{Hojjatpanah2019}; \citealt{Arentsen2019}). Metal-rich stars with low $Z$ distances and retrograde movements have been identified as members of the in-situ formed halo, which has a wide [Fe/H] distribution that ranges from $-2$ dex to the solar value according to the analysis of \citet{bonaca2017ApJ...845..101B}.
There, the authors discuss that radial migration, promoted by the accretion of intergalactic medium and satellite galaxies, is a mechanism that can possibly explain in-situ halo stars well dispersed around $V = 0\ \text{kms}^{-1}$ in the Toomre diagram. 
In such a case, HIP 24186 should be older than the rest of M dwarfs in the current sample to be consistent with a metal-rich halo formed prior to the thin disc \citep{2017MNRAS.467.2430M}.
Still, although it is rare, a few in-situ halo stars may invade the area covered by the discs as shown by the simulations presented in \citet{bonaca2017ApJ...845..101B}.

Metal-rich stars with halo kinematics can be also members of the  so called "splashed disc" structure, the dynamics of which is interpreted as the consequence of the accretion events. According to \citet{El-Badry2016ApJ...820..131E}, some stars can be formed during gas outflows driven by stellar feedback, thus they may have quite eccentric initial orbits. In such a case, HIP 24186 could be dated contemporary to the merging event that promoted its formation, hence it should be somewhat younger than in-situ halo scattered stars.

Therefore, our [Fe/H] data plus kinematics strongly suggest that our sample is fully dominated by thin and thick disk stars, as expected.

\section{Conclusions}
\label{sec7}
We present T$_\text{eff}$ and [Fe/H] determinations for 178 M dwarfs located within 25 parsecs of the Sun, by means of a new system of 147 spectral indices measured in the $\lambda \lambda$ 8390-8834 \AA\ range and coupled to the $J-H$ color. The indices are defined in moderate resolution (R $\sim$ 11,000) and moderately high S/N spectra ($\gtrsim$ 100), obtained at the coudé spectrograph of the 1.60 m Perkin-Elmer telescope of Observatório do Pico dos Dias, Brazil. A PCA regression applied to the indices and $J-H$ color was calibrated for 44 M dwarfs with well determined parameters (\citealt{Mann2015}). The calibration achieved, respectively, an internal precision of 81 K and 0.12 dex for T$_\text{eff}$ and [Fe/H]. Total median uncertainties estimated from Monte Carlo propagation of errors are, respectively, 105 K and 0.23 dex for T$_\text{eff}$ and [Fe/H]. Radial velocities with mean internal precision of 1.4 km/s are also presented, for many objects for the first time.

Our conclusions are as follows:
\begin{itemize}
    \item the comparison of our atmospheric parameters with others available in the literature, employing a wide variety of both photometric and spectroscopic methods, reveals considerable discrepancies between published works, both in zero-point and scale, up to a few hundred Kelvin in the worst cases. The median differences of 75 $\pm$ 273 K and 0.02 $\pm$ 0.31 dex for T$_\text{eff}$ and [Fe/H], respectively. Our values are thus accurate but present lower precision when compared to those of most authors;
    \item the PCA-calibrated spectral indices constitute a model-free, rather competitive technique able to retrieve atmospheric parameters with good accuracy and reasonable precision, well suited to the extensive databases of M dwarf spectra in the far red;
    \item our raw, uncorrected metallicity distribution function of [Fe/H] for nearby M dwarfs shows a peak at [Fe/H] $\sim$ -0.10 dex, in very good agreement with the RAVE (\citealt{Boeche2013}) metallicity distribution function for 19,962 FGK dwarf stars, most of them contained within 0.0 kpc $<\left|\text{Z}_{\text{max}}\right| <$ 0.4 kpc, corresponding to the local thin disk. The MDFs for FGKM spectral types appear thus to be consistent to one another;
    \item investigation of the Galactic orbits of the sample stars shows that the whole sample can be safely ascribed to the thin and thick disk, the only notable exception being the well-know metal-poor red dwarf Kapteyn's star. Our modelling strong suggests the complete dominance of thick and thin disk stars in our sample.
\end{itemize}

Current and future projects focusing in M dwarfs (such as CARMENES, \citealt{Carmenes}) can benefit from our spectral index approach which is able to derive accurate T$_\text{eff}$ and [Fe/H] for large samples of M dwarfs employing spectra of relatively modest quality. Gaia Data Release 3 Catalogue (expected to the first half of 2022) will contain spectra of millions of stars -- spectra which will have a similar resolution to the ones used in this work. Thus, our method can me easily applied to the millions of M dwarfs that Gaia contemplates. This represents an unique opportunity to globally characterize the most abundant objects in the Galaxy.

\section*{Data availability}
The data underlying this article will be shared on reasonable request to the corresponding author.

\section*{Acknowledgements}
E.C.A. acknowledges CNPq/Brazil and CAPES/Brazil scholarships. G.F.P.M. acknowledges grant 474972/2009-7 from CNPq/Brazil. M.L.U.M. acknowledges CAPES/Brazil and FAPERJ scholarships. D.L.O. acknowledges a CAPES/Brazil scholarship and a FAPESP 2016/20667-8 grant. R.E.G. acknowledges scholarships from CAPES/Brazil and ESO, and the support by the National Science Centre, Poland, through project 2018/31/B/ST9/01469. We thank H. J. Rocha-Pinto for helpful discussions. We thank the referee, Dr. Eduardo Martin, for criticism and suggestions that considerably improved the manuscript. We thank the staff of OPD/LNA for considerable support in the many observing runs carried out during this project. Use was made of the Simbad database, operated at the CDS, Strasbourg, France, and of NASA’s Astrophysics Data System Bibliographic Services. This publication makes use of data products from the Two Micron All Sky Survey, which is a joint project of the University of Massachusetts and the Infrared Processing and Analysis Center/California Institute of Technology, funded by the National Aeronautics and Space Administration and the National Science Foundation. This work presents results from the European Space Agency (ESA) space mission Gaia. The Gaia data are being processed by the Gaia Data Processing and Analysis Consortium (DPAC). Funding for the DPAC is provided by national institutions, in particular the institutions participating in the Gaia MultiLateral Agreement (MLA). The Gaia mission website is \url{https://www.cosmos.esa.int/gaia}. The Gaia archive website is \url{https://archives.esac.esa.int/gaia}.
This research made use of Astropy\footnote{http://www.astropy.org}, a community-developed core Python package for Astronomy \citep{astropy2013A&A...558A..33A, astropy:2018}.




\bibliographystyle{mnras}
\bibliography{references} 




\appendix

\section{Data}
\begin{table*}
\scriptsize
\caption{Spectral indices defined from 8390.170 to 8834.470 \AA\ with final percentual internal relative uncertainties. We indicate with ``$\times$" symbol the indices that were not used in the PCA regression. The full version is available as supplementary material.}
\begin{tabular}{lccc|lccc|lccc}
\hline \hline 
ID	&	$\mathrm{\lambda}_\text{i}$	&	$\mathrm{\lambda}_\text{f}$ & $\sigma^{\text{rel}}_{\text{final}}$ \Tstrut & ID	&	$\mathrm{\lambda}_\text{i}$	&	$\mathrm{\lambda}_\text{f}$ & $\sigma^{\text{rel}}_{\text{final}}$ \Tstrut & ID	&	$\mathrm{\lambda}_\text{i}$	&	$\mathrm{\lambda}_\text{f}$ & $\sigma^{\text{rel}}_{\text{final}}$ \Tstrut\\
	& (\AA)	& (\AA) & 	& (\AA)	& (\AA) & 	& (\AA)	& (\AA) &\\ 
\hline
\hline
$i1^{\times}$ & 8390.170 & 8392.852 & 0.20 &	$i76$ & 8579.369 & 8581.406 & 0.15 &	$i151$ & 8774.903 & 8777.156 & 0.50 \\
$i2^{\times}$ & 8392.852 & 8395.796 & 0.14 &	$i77$ & 8581.406 & 8582.777 & 0.11 &	$i152$ & 8777.156 & 8779.373 & 0.18 \\
$i3^{\times}$ & 8395.796 & 8399.997 & 0.07 &	$i78$ & 8582.777 & 8583.871 & 0.13 &	$i153$ & 8779.373 & 8782.086 & 0.28 \\
$i4^{\times}$ & 8399.997 & 8401.949 & 0.12 &	$i79$ & 8583.871 & 8585.048 & 0.21 &	$i154$ & 8782.086 & 8783.529 & 0.27 \\
$i5^{\times}$ & 8401.949 & 8405.622 & 0.18 &	$i80$ & 8585.048 & 8587.833 & 0.20 &	$i155$ & 8783.529 & 8785.283 & 0.30 \\
$i6^{\times}$ & 8405.622 & 8408.070 & 0.23 &	$i81$ & 8587.833 & 8589.910 & 0.21 &	$i156$ & 8785.283 & 8788.175 & 0.29 \\
$i7^{\times}$ & 8408.070 & 8411.035 & 0.17 &	$i82$ & 8589.910 & 8595.833 & 0.18 &	$i157$ & 8788.175 & 8791.172 & 0.30 \\
$i8^{\times}$ & 8411.035 & 8413.815 & 0.05 &	$i83$ & 8595.833 & 8597.521 & 0.19 &	$i158$ & 8791.172 & 8795.837 & 0.18 \\
$i9^{\times}$ & 8413.815 & 8415.469 & 0.28 &	$i84$ & 8597.521 & 8600.498 & 0.14 &	$i159$ & 8795.837 & 8799.763 & 0.39 \\
$i10^{\times}$ & 8415.469 & 8418.413 & 0.12 &	$i85$ & 8600.498 & 8601.953 & 0.22 &	$i160$ & 8799.763 & 8802.237 & 0.10 \\
$i11^{\times}$ & 8418.413 & 8421.854 & 0.13 &	$i86$ & 8601.953 & 8604.931 & 0.21 &	$i161$ & 8802.237 & 8803.944 & 0.11 \\
$i12^{\times}$ & 8421.854 & 8423.765 & 0.10 &	$i87$ & 8604.931 & 8607.399 & 0.23 &	$i162$ & 8803.944 & 8805.366 & 0.10 \\
$i13^{\times}$ & 8423.765 & 8425.526 & 0.13 &	$i88$ & 8607.399 & 8609.319 & 0.26 &	$i163$ & 8805.366 & 8808.375 & 0.05 \\
$i14^{\times}$ & 8425.526 & 8427.743 & 0.03 &	$i89$ & 8609.319 & 8610.403 & 0.19 &	$i164$ & 8808.375 & 8812.309 & 0.18 \\
$i15^{\times}$ & 8427.743 & 8429.463 & 0.18 &	$i90$ & 8610.403 & 8613.631 & 0.09 &	$i165$ & 8812.309 & 8817.409 & 0.28 \\
$i16^{\times}$ & 8429.463 & 8430.940 & 0.20 &	$i91$ & 8613.631 & 8618.739 & 0.14 &	$i166$ & 8817.409 & 8820.584 & 0.25 \\
$i17^{\times}$ & 8430.940 & 8432.396 & 0.21 &	$i92$ & 8618.739 & 8620.138 & 0.23 &	$i167$ & 8820.584 & 8823.114 & 0.12 \\
$i18^{\times}$ & 8432.396 & 8436.829 & 0.02 &	$i93$ & 8620.138 & 8622.354 & 0.14 &	$i168$ & 8823.114 & 8826.059 & 0.05 \\
$i19^{\times}$ & 8436.829 & 8440.270 & 0.06 &	$i94$ & 8622.354 & 8627.392 & 0.20 &	$i169$ & 8826.059 & 8830.919 & 0.15 \\
$i20$ & 8440.270 & 8445.199 & 0.09 &	$i95$ & 8627.392 & 8629.836 & 0.20 &	$i170$ & 8830.919 & 8834.470 & 0.26 \\
$i21$ & 8445.199 & 8449.885 & 0.10 &	$i96$ & 8629.836 & 8631.937 & 0.22 \\	
$i22$ & 8449.885 & 8451.508 & 0.07 &	$i97$ & 8631.937 & 8635.377 & 0.16 \\	
$i23$ & 8451.508 & 8454.517 & 0.07 &	$i98$ & 8635.377 & 8638.090 & 0.23 \\	
$i24$ & 8454.517 & 8456.237 & 0.06 &	$i99$ & 8638.090 & 8640.34 & 0.21 \\	
$i25$ & 8456.237 & 8458.719 & 0.05 &	$i100$ & 8640.34 & 8643.583 & 0.21 \\	
$i26$ & 8458.719 & 8461.663 & 0.08 &	$i101$ & 8643.583 & 8646.943 & 0.24 \\	
$i27$ & 8461.663 & 8466.315 & 0.11 &	$i102$ & 8646.943 & 8649.393 & 0.19 \\	
$i28$ & 8466.315 & 8471.245 & 0.05 &	$i103$ & 8649.393 & 8653.064 & 0.17 \\	
$i29$ & 8471.245 & 8473.693 & 0.08 &	$i104$ & 8653.064 & 8658.494 & 0.15 \\	
$i30$ & 8473.693 & 8475.182 & 0.08 &	$i105^{\times}$ & 8658.494 & 8666.718 & 0.05 \\	
$i31$ & 8475.182 & 8477.896 & 0.11 &	$i106$ & 8666.718 & 8669.528 & 0.15 \\	
$i32$ & 8477.896 & 8479.086 & 0.10 &	$i107$ & 8669.528 & 8671.017 & 0.23 \\	
$i33$ & 8479.086 & 8480.575 & 0.10 &	$i108$ & 8671.017 & 8672.241 & 0.19 \\	
$i34$ & 8480.575 & 8482.030 & 0.09 &	$i109$ & 8672.241 & 8673.962 & 0.22 \\	
$i35$ & 8482.030 & 8483.751 & 0.09 &	$i110$ & 8673.962 & 8676.410 & 0.06 \\	
$i36$ & 8483.751 & 8485.008 & 0.11 &	$i111$ & 8676.410 & 8677.634 & 0.31 \\	
$i37$ & 8485.008 & 8486.265 & 0.11 &	$i112$ & 8677.634 & 8680.372 & 0.26 \\	
$i38$ & 8486.265 & 8489.429 & 0.11 &	$i113$ & 8680.372 & 8682.300 & 0.25 \\	
$i39$ & 8489.429 & 8492.109 & 0.13 &	$i114$ & 8682.300 & 8685.496 & 0.14 \\	
$i40$ & 8492.109 & 8494.821 & 0.10 &	$i115$ & 8685.496 & 8687.480 & 0.15 \\	
$i41$ & 8494.821 & 8496.311 & 0.06 &	$i116$ & 8687.480 & 8690.203 & 0.05 \\	
$i42^{\times}$ & 8496.311 & 8499.288 & 0.03 &	$i117$ & 8690.203 & 8691.418 & 0.17 \\	
$i43$ & 8499.288 & 8501.703 & 0.07 &	$i118$ & 8691.418 & 8693.370 & 0.10 \\	
$i44$ & 8501.703 & 8503.266 & 0.10 &	$i119$ & 8693.370 & 8694.594 & 0.16 \\	
$i45$ & 8503.266 & 8508.076 & 0.09 &	$i120$ & 8694.594 & 8695.818 & 0.14 \\	
$i46$ & 8508.076 & 8509.565 & 0.12 &	$i121$ & 8695.818 & 8696.847 & 0.18 \\	
$i47$ & 8509.565 & 8511.285 & 0.11 &	$i122$ & 8696.847 & 8698.531 & 0.16 \\	
$i48$ & 8511.285 & 8513.271 & 0.11 &	$i123$ & 8698.531 & 8700.482 & 0.12 \\	
$i49$ & 8513.271 & 8514.726 & 0.03 &	$i124$ & 8700.482 & 8702.965 & 0.18 \\	
$i50$ & 8514.726 & 8517.083 & 0.08 &	$i125$ & 8702.965 & 8704.903 & 0.19 \\	
$i51$ & 8517.083 & 8522.137 & 0.07 &	$i126$ & 8704.903 & 8707.893 & 0.19 \\	
$i52$ & 8522.137 & 8523.361 & 0.09 &	$i127$ & 8707.893 & 8709.337 & 0.20 \\	
$i53$ & 8523.361 & 8525.313 & 0.10 &	$i128$ & 8709.337 & 8712.546 & 0.11 \\	
$i54$ & 8525.313 & 8527.034 & 0.09 &	$i129$ & 8712.546 & 8715.227 & 0.13 \\	
$i55$ & 8527.034 & 8529.238 & 0.11 &	$i130$ & 8715.227 & 8716.715 & 0.13 \\	
$i56$ & 8529.238 & 8530.958 & 0.12 &	$i131$ & 8716.715 & 8718.171 & 0.13 \\	
$i57$ & 8530.958 & 8533.142 & 0.10 &	$i132$ & 8718.171 & 8720.421 & 0.15 \\	
$i58$ & 8533.142 & 8535.855 & 0.09 &	$i133$ & 8720.421 & 8723.101 & 0.18 \\	
$i59$ & 8535.855 & 8539.563 & 0.08 &	$i134$ & 8723.101 & 8726.066 & 0.16 \\	
$i60^{\times}$ & 8539.563 & 8543.961 & 0.03 &	$i135$ & 8726.066 & 8728.514 & 0.21 \\	
$i61$ & 8543.961 & 8545.945 & 0.05 &	$i136$ & 8728.514 & 8732.683 & 0.16 \\	
$i62$ & 8545.945 & 8549.374 & 0.07 &	$i137$ & 8732.683 & 8736.854 & 0.11 \\	
$i63$ & 8549.374 & 8551.095 & 0.12 &	$i138$ & 8736.854 & 8739.796 & 0.14 \\	
$i64$ & 8551.095 & 8553.543 & 0.13 &	$i139$ & 8739.796 & 8742.277 & 0.18 \\	
$i65$ & 8553.543 & 8556.365 & 0.14 &	$i140$ & 8742.277 & 8744.713 & 0.22 \\	
$i66$ & 8556.365 & 8557.745 & 0.16 &	$i141$ & 8744.713 & 8746.201 & 0.23 \\	
$i67$ & 8557.745 & 8561.681 & 0.13 &	$i142$ & 8746.201 & 8748.636 & 0.13 \\	
$i68$ & 8561.681 & 8563.634 & 0.14 &	$i143$ & 8748.636 & 8752.091 & 0.16 \\	
$i69$ & 8563.634 & 8565.089 & 0.18 &	$i144$ & 8752.091 & 8754.075 & 0.24 \\	
$i70$ & 8565.089 & 8569.047 & 0.17 &	$i145$ & 8754.075 & 8757.979 & 0.14 \\	
$i71$ & 8569.047 & 8572.487 & 0.14 &	$i146$ & 8757.979 & 8759.700 & 0.32 \\	
$i72$ & 8572.487 & 8574.048 & 0.17 &	$i147$ & 8759.700 & 8764.617 & 0.17 \\	
$i73$ & 8574.048 & 8575.135 & 0.15 &	$i148$ & 8764.617 & 8769.051 & 0.15 \\	
$i74$ & 8575.135 & 8577.385 & 0.18 &	$i149$ & 8769.051 & 8770.989 & 0.33 \\	
$i75$ & 8577.385 & 8579.369 & 0.20 &	$i150$ & 8770.989 & 8774.903 & 0.11 \\	

\end{tabular}
\label{tab:appendix_indices}
\end{table*}
\begin{table*}
    \scriptsize
    \centering
    \caption{Column 1 lists the variables, i.e., color and spectral indices. Columns 2 and 3 list the mean and standard deviation used to normalize the variables in the form $\text{Variable}_j^\text{norm} = (\text{Variable}_j-\text{Mean}_j)/\text{Std}_j$ -- in order to get a variables' distribution with 0 and 1 mean and standard deviation, respectively. Columns 4, 5, 6 and 7 list the coefficients derived by the PCA regression that, being multiplied by the normalized variables' vector, create the Principal Components. The full version is available as supplementary material.}
        \begin{tabular}{ccccccc}
        \hline
        \hline
        Variable & Mean & Std & PC1 & PC2 & PC3 & PC5 \\
        \hline
        $J-H$ & 0.59443182 & 0.04552709 & 0.017392992 & -0.042049815 & 0.487154363 & 0.3637667876 \\
        $i$20 & 0.40671511 & 0.25811607 & -0.082275196 & 0.102241666 & -0.0112641841 & -0.0396752735 \\
        $i$21 & 0.34567894 & 0.23123022 & -0.082916601 & 0.091045205 & 0.0219376739 & -0.0431499682 \\
        $i$22 & 0.16982121 & 0.06566573 & -0.081422918 & 0.09004488 & 0.0974459883 & -0.0275029256 \\
        $i$23 & 0.33903625 & 0.22863852 & -0.08016789 & 0.129052651 & -0.001602449 & -0.0585099458 \\
        $i$24 & 0.17617413 & 0.11276828 & -0.080625134 & 0.123963189 & 0.0268581814 & -0.0767561055 \\
        $i$25 & 0.28391504 & 0.16430488 & -0.081562147 & 0.114327971 & 0.0205351021 & -0.03033259 \\
        $i$26 & 0.2766045 & 0.18470607 & -0.082831315 & 0.097593103 & -0.0156747225 & -0.0181320588 \\
        $i$27 & 0.43927773 & 0.29521656 & -0.081657078 & 0.107551106 & -0.0069125349 & -0.0797381373 \\
        $i$28 & 0.73391553 & 0.29109171 & -0.083897916 & 0.074563686 & 0.0331092506 & -0.0454053418 \\
        $i$29 & 0.28032803 & 0.18112735 & -0.082024002 & 0.107233962 & -0.0142134392 & -0.0176886425 \\
        $i$30 & 0.14192388 & 0.10282675 & -0.082282485 & 0.105699632 & 0.0040490258 & -0.0274885583 \\
        $i$31 & 0.25109636 & 0.18126266 & -0.082574778 & 0.103206163 & 0.0033750912 & -0.0043288407 \\
        $i$32 & 0.11171872 & 0.07570451 & -0.082989697 & 0.09634329 & 0.0173748786 & -0.0293256437 \\
        $i$33 & 0.13897736 & 0.09498492 & -0.082386003 & 0.103695897 & 0.037052364 & -0.0318980012 \\
        $i$34 & 0.14980371 & 0.0956114 & -0.082305235 & 0.100722106 & 0.0326366776 & -0.0625237795 \\
        $i$35 & 0.17474708 & 0.10675363 & -0.082510201 & 0.099295606 & 0.040152503 & -0.0817328513 \\
        $i$36 & 0.10638822 & 0.07583941 & -0.082097577 & 0.107864225 & 0.0130254591 & -0.0592137083 \\
        $i$37 & 0.1156355 & 0.07917762 & -0.082569833 & 0.100782138 & -0.0032514927 & -0.0348488209 \\
        $i$38 & 0.28822427 & 0.1922289 & -0.083776641 & 0.084411983 & -0.0057537133 & -0.0567331916 \\
        $i$39 & 0.2207851 & 0.15017105 & -0.084171267 & 0.07352147 & -0.0159738402 & -0.0358892318 \\
        $i$40 & 0.26158848 & 0.14448136 & -0.083649013 & 0.06086956 & 0.0017600052 & -0.0636953339 \\
        $i$41 & 0.2106442 & 0.07901366 & -0.08425531 & 0.054744858 & 0.0648456553 & -0.0071138349 \\
        $i$43 & 0.27783231 & 0.12258709 & -0.085013136 & 0.051253744 & 0.0320752936 & -0.0191469705 \\
        $i$44 & 0.15309916 & 0.08841463 & -0.084603989 & 0.061560708 & 0.022897678 & -0.0662439206 \\
        $i$45 & 0.5406636 & 0.34914814 & -0.083736937 & 0.084893685 & -0.0057156965 & -0.0334389763 \\
        $i$46 & 0.14953586 & 0.1084143 & -0.083462408 & 0.089172534 & -0.0167264217 & 0.0108811494 \\
        $i$47 & 0.17242562 & 0.11910837 & -0.083636857 & 0.077447803 & -0.0001118251 & 0.0346563557 \\
        $i$48 & 0.20924455 & 0.12814084 & -0.083538627 & 0.072587657 & 0.0382135558 & -0.027667984 \\
        $i$49 & 0.29767027 & 0.0671156 & -0.081448631 & 0.039347755 & 0.1675914636 & -0.1192314997 \\
        $i$50 & 0.32637413 & 0.16502801 & -0.082371403 & 0.100934161 & 0.0663593605 & -0.035202341 \\
        $i$51 & 0.6736822 & 0.33799078 & -0.082831609 & 0.098841273 & 0.0474689276 & -0.0363919954 \\
        $i$52 & 0.13360598 & 0.08657209 & -0.082013961 & 0.111101215 & 0.0202782809 & -0.0422074209 \\
        $i$53 & 0.19331083 & 0.12564916 & -0.083007992 & 0.098445338 & 0.0303926288 & -0.0307451552 \\
        $i$54 & 0.19902754 & 0.10392635 & -0.082987935 & 0.082269989 & 0.0938319146 & -0.0333006905 \\
        $i$55 & 0.22834189 & 0.13862311 & -0.083334072 & 0.092674914 & 0.0204756515 & -0.0439372845 \\
        $i$56 & 0.18343598 & 0.10958116 & -0.083385953 & 0.092120643 & 0.0203166122 & -0.0335679888 \\
        $i$57 & 0.23967167 & 0.13125559 & -0.083917417 & 0.080302744 & 0.0332116875 & -0.0240026929 \\
        $i$58 & 0.30578773 & 0.14822983 & -0.084651265 & 0.061866683 & 0.0205812602 & -0.0133720491 \\
        $i$59 & 0.56545909 & 0.14626777 & -0.083609819 & -0.004036885 & 0.1059674321 & -0.0414764408 \\
        $i$61 & 0.35938996 & 0.0594687 & -0.076747366 & -0.064698363 & 0.1779508299 & -0.0213651654 \\
        $i$62 & 0.4666589 & 0.13878976 & -0.084022835 & -0.005341379 & 0.0794070287 & -0.0058885972 \\
        $i$63 & 0.17247625 & 0.08503004 & -0.085408022 & 0.025239784 & 0.0112918046 & -0.0058620395 \\
        $i$64 & 0.22881712 & 0.13116281 & -0.085196157 & 0.039568153 & -0.020046792 & 0.0163914135 \\
        $i$65 & 0.25810515 & 0.1503115 & -0.085683392 & 0.030950981 & 0.0072631479 & -0.000230114 \\
        $i$66 & 0.11290471 & 0.07085985 & -0.085125589 & 0.02937078 & 0.0194240275 & 0.0118715837 \\
        $i$67 & 0.42480409 & 0.26700952 & -0.085376442 & 0.053392493 & -0.017039493 & -0.0117187421 \\
        $i$68 & 0.18615992 & 0.12043932 & -0.085484332 & 0.0481424 & -0.0098387536 & -0.0073821227 \\
        $i$69 & 0.12522229 & 0.08459503 & -0.08529374 & 0.048980084 & -0.0194800051 & 0.0035443405 \\
        $i$70 & 0.3313336 & 0.2225693 & -0.085587777 & 0.037453701 & -0.0142085857 & 0.0239196918 \\
        $i$71 & 0.337685 & 0.21149735 & -0.085443237 & 0.048544178 & -0.0072350343 & 0.0221100044 \\
        $i$72 & 0.14513782 & 0.09656165 & -0.085299948 & 0.047476406 & -0.025310022 & 0.0158300183 \\
        $i$73 & 0.1007517 & 0.06500711 & -0.085681466 & 0.03270413 & 0.0039079475 & 0.0216945445 \\
        $i$74 & 0.18967881 & 0.12790317 & -0.085540173 & 0.03238452 & -0.0229149245 & 0.0267455125 \\
        $i$75 & 0.16195949 & 0.10910287 & -0.085578265 & 0.035349095 & -0.0129149776 & 0.0371118815 \\
        $i$76 & 0.1805003 & 0.12677026 & -0.085447085 & 0.046934136 & -0.026258472 & 0.0157692041 \\
        $i$77 & 0.17198193 & 0.06949543 & -0.085052113 & 0.003178242 & 0.0913985651 & 0.05774963 \\
        $i$78 & 0.11141617 & 0.05873708 & -0.085150862 & 0.03029419 & 0.0483528909 & 0.0638639443 \\
        $i$79 & 0.10197177 & 0.07042437 & -0.085199173 & 0.04578849 & -0.0134546543 & 0.052824858 \\
        $i$80 & 0.22011418 & 0.15613495 & -0.085321085 & 0.043226164 & -0.0191665521 & 0.0203567645 \\
        $i$81 & 0.16815323 & 0.11673784 & -0.085381178 & 0.03367642 & -0.0214518198 & 0.0636497568 \\
        $i$82 & 0.51671208 & 0.3461347 & -0.08539129 & 0.035407955 & -0.0123373715 & 0.0513233678 \\
        $i$83 & 0.13415504 & 0.09485464 & -0.085209877 & 0.033157009 & -0.0066489261 & 0.086600965 \\
        $i$84 & 0.25111545 & 0.15055119 & -0.085406906 & 0.017561489 & 0.0302995219 & 0.087476817 \\
        $i$85 & 0.11461999 & 0.07909896 & -0.085278047 & 0.029268181 & -0.0120657492 & 0.0585266968 \\
        $i$86 & 0.21697517 & 0.15835267 & -0.085508824 & 0.027742012 & -0.0011432453 & 0.0806874066 \\
        $i$87 & 0.17528893 & 0.13133744 & -0.085690722 & 0.020736893 & -0.0139773827 & 0.082355655 \\
        $i$88 & 0.12458028 & 0.09819802 & -0.085466329 & 0.021322793 & -0.0056088681 & 0.0982505503 \\
        $i$89 & 0.07009203 & 0.05384342 & -0.085373352 & 0.02033012 & 0.0311285425 & 0.056616093 \\
        $i$90 & 0.35107481 & 0.155991 & -0.08592485 & -0.001378498 & 0.0490261307 & 0.038507224 \\
        $i$91 & 0.40714337 & 0.27731662 & -0.086304055 & 0.003863258 & -0.0019279218 & 0.0096802999 \\
        $i$92 & 0.09819069 & 0.07729978 & -0.085938524 & 0.004431976 & -0.0194661584 & 0.0694594886 \\
        $i$93 & 0.18691186 & 0.09780898 & -0.085058813 & -0.022062821 & 0.064659784 & 0.0785997954 \\
        $i$94 & 0.34259916 & 0.26389809 & -0.085779871 & 0.025161904 & 0.0149124415 & 0.036730694 \\
        $i$95 & 0.1741404 & 0.13177024 & -0.085303701 & 0.034895397 & 0.0154393849 & 0.0141977337 \\
        \hline
        \end{tabular}
\label{tab:PCA}
\end{table*}

\begin{table*}
    \scriptsize
    \centering
        \begin{tabular}{ccccccc}
        \hline
        \hline
        Variable & Mean & Std & PC1 & PC2 & PC3 & PC5 \\
        \hline
        $i$96 & 0.14986292 & 0.11239392 & -0.085530254 & 0.038687116 & 0.0236438043 & 0.0015332313 \\
        $i$97 & 0.25146989 & 0.15548816 & -0.084746459 & 0.017098309 & 0.0735126637 & 0.0529481037 \\
        $i$98 & 0.16671691 & 0.12586519 & -0.084801798 & 0.033993351 & 0.050251354 & 0.0560871347 \\
        $i$99 & 0.16163798 & 0.1212219 & -0.08511923 & 0.037569575 & 0.0170764249 & 0.002045354 \\
        $i$100 & 0.19308294 & 0.14574406 & -0.0841996 & 0.030821975 & 0.0616532279 & 0.0850513918 \\
        $i$101 & 0.24536052 & 0.17649419 & -0.084808437 & 0.044098522 & 0.0129914758 & 0.0557515932 \\
        $i$102 & 0.17845078 & 0.11989134 & -0.084964086 & 0.034510739 & 0.0593092182 & 0.0269608975 \\
        $i$103 & 0.30263447 & 0.17886801 & -0.085030855 & 0.033735714 & 0.0372890386 & 0.0308712062 \\
        $i$104 & 0.48917443 & 0.23093181 & -0.084847443 & -0.046249202 & 0.0110871613 & 0.082376765 \\
        $i$106 & 0.25341837 & 0.1199623 & -0.083846875 & -0.057183794 & -0.0051843858 & 0.1057328852 \\
        $i$107 & 0.10881087 & 0.06487943 & -0.08338185 & -0.058387588 & -0.0187403112 & 0.1134804373 \\
        $i$108 & 0.09347712 & 0.05517833 & -0.08385096 & -0.059115886 & -0.0001335588 & 0.0960615466 \\
        $i$109 & 0.13286617 & 0.08000828 & -0.084566686 & -0.048754265 & -0.0132525346 & 0.0994285892 \\
        $i$110 & 0.39899034 & 0.09475892 & -0.079051851 & -0.118830695 & 0.0569024501 & 0.0929194081 \\
        $i$111 & 0.08174456 & 0.05610179 & -0.083883088 & -0.056415405 & -0.0659383089 & 0.0894565046 \\
        $i$112 & 0.1917222 & 0.12320819 & -0.084421395 & -0.046238828 & -0.0128747296 & 0.1108493317 \\
        $i$113 & 0.14497502 & 0.09163848 & -0.084909902 & -0.045628514 & -0.0395547632 & 0.0872115705 \\
        $i$114 & 0.32554981 & 0.13853283 & -0.083561118 & -0.078211888 & -0.0322393115 & 0.0708668027 \\
        $i$115 & 0.17619167 & 0.08793204 & -0.082343702 & -0.09577231 & 0.0127919372 & 0.0827291882 \\
        $i$116 & 0.62655852 & 0.09989846 & -0.068484376 & -0.18107765 & 0.1731441551 & -0.0125561895 \\
        $i$117 & 0.11495466 & 0.06140391 & -0.083424169 & -0.070461809 & -0.0162091662 & 0.0255055757 \\
        $i$118 & 0.25455739 & 0.09924167 & -0.083782617 & -0.046954947 & 0.0003919154 & -0.0198085932 \\
        $i$119 & 0.12906973 & 0.07324856 & -0.084777769 & 0.001276274 & -0.0177803436 & 0.0051510911 \\
        $i$120 & 0.13330996 & 0.07231874 & -0.084272988 & 0.001289059 & -0.0160397648 & -0.0402312173 \\
        $i$121 & 0.08700899 & 0.05544443 & -0.085000738 & 0.001366778 & 0.0033414296 & -0.0058640567 \\
        $i$122 & 0.15875174 & 0.09010921 & -0.084127665 & -0.004457592 & 0.0148194225 & -0.0579989841 \\
        $i$123 & 0.17666765 & 0.08926536 & -0.083563734 & -0.034959217 & 0.0945905172 & -0.0002001871 \\
        $i$124 & 0.20206178 & 0.12420656 & -0.085025307 & -0.013641296 & 0.0106194457 & -0.0024370258 \\
        $i$125 & 0.13402992 & 0.08834963 & -0.085121241 & -0.032991536 & -0.001239109 & 0.0138052988 \\
        $i$126 & 0.22038208 & 0.14499373 & -0.085386265 & -0.022426864 & -0.0096335616 & 0.0287389188 \\
        $i$127 & 0.09458858 & 0.06719556 & -0.084079784 & -0.037819577 & -0.014904956 & 0.0355566893 \\
        $i$128 & 0.3222128 & 0.17578936 & -0.083293241 & -0.047914507 & 0.0150744289 & -0.0362613132 \\
        $i$129 & 0.25371985 & 0.13538776 & -0.083534554 & -0.050524035 & 0.0263859421 & -0.0077108058 \\
        $i$130 & 0.14098875 & 0.08167171 & -0.084359503 & -0.015926074 & -0.0242651826 & -0.027729165 \\
        $i$131 & 0.13837667 & 0.07012221 & -0.084320812 & -0.03861794 & 0.0280314692 & -0.0450300601 \\
        $i$132 & 0.20547015 & 0.11691041 & -0.083825194 & -0.037601267 & -0.0080331835 & -0.0279041358 \\
        $i$133 & 0.2035689 & 0.13954051 & -0.084553022 & -0.026431344 & -0.0435087342 & 0.0102356766 \\
        $i$134 & 0.22630705 & 0.14891604 & -0.08439553 & -0.019851047 & -0.015080093 & -0.0331918392 \\
        $i$135 & 0.14485144 & 0.10269643 & -0.083858716 & -0.027535519 & 0.0105171787 & -0.0063489142 \\
        $i$136 & 0.32670534 & 0.19054708 & -0.084644459 & -0.034204075 & -0.0105706093 & -0.0530043479 \\
        $i$137 & 0.42440284 & 0.18238542 & -0.083550553 & -0.084132632 & 0.0181960073 & -0.0672251183 \\
        $i$138 & 0.2666453 & 0.150199 & -0.08468076 & -0.052538734 & -0.0544165006 & -0.0556925672 \\
        $i$139 & 0.19195314 & 0.10833321 & -0.083481147 & -0.082511863 & -0.0361135984 & -0.0778330341 \\
        $i$140 & 0.14168004 & 0.09815464 & -0.082294754 & -0.096145876 & -0.0786394827 & -0.020821101 \\
        $i$141 & 0.09165217 & 0.06351014 & -0.082325465 & -0.084849193 & -0.0799766491 & -0.0216610276 \\
        $i$142 & 0.22090064 & 0.11881728 & -0.083492197 & -0.075933668 & -0.055982219 & -0.0499344981 \\
        $i$143 & 0.2632828 & 0.16096411 & -0.08378071 & -0.077209399 & -0.0661129701 & -0.0537634925 \\
        $i$144 & 0.12137028 & 0.08264403 & -0.080940723 & -0.108901315 & -0.0900960891 & -0.0518953711 \\
        $i$145 & 0.36233049 & 0.16467008 & -0.081295782 & -0.110222972 & -0.0509945764 & -0.0519368158 \\
        $i$146 & 0.08914297 & 0.0672808 & -0.080436193 & -0.106851794 & -0.1079336937 & -0.0547042879 \\
        $i$147 & 0.3239767 & 0.17990019 & -0.079248737 & -0.135378838 & -0.0520803502 & -0.0370343542 \\
        $i$148 & 0.35987273 & 0.19299453 & -0.081134383 & -0.078152306 & -0.0907787291 & -0.1136752872 \\
        $i$149 & 0.09972542 & 0.07369268 & -0.079745333 & -0.097967783 & -0.1004904367 & -0.0383242451 \\
        $i$150 & 0.36269583 & 0.10405323 & -0.060609417 & -0.221361588 & 0.0958686969 & 0.0399223204 \\
        $i$151 & 0.0704919 & 0.06600187 & -0.067887449 & -0.149690929 & -0.1435433236 & -0.0253037874 \\
        $i$152 & 0.19129848 & 0.09742386 & -0.080517733 & -0.078841029 & -0.0795107163 & -0.1389866999 \\
        $i$153 & 0.17840902 & 0.11412152 & -0.081143341 & -0.092831866 & -0.1041369486 & -0.0989746055 \\
        $i$154 & 0.09120272 & 0.05949992 & -0.079309862 & -0.106387255 & -0.1338970962 & -0.1194024543 \\
        $i$155 & 0.09642644 & 0.06441347 & -0.077330977 & -0.138100508 & -0.0910398564 & -0.045581462 \\
        $i$156 & 0.15225897 & 0.10992801 & -0.078475988 & -0.109140151 & -0.1417457408 & -0.1138915413 \\
        $i$157 & 0.15272985 & 0.09950162 & -0.075569351 & -0.14987268 & -0.0855162457 & -0.1271282305 \\
        $i$158 & 0.28520705 & 0.15322724 & -0.073565773 & -0.175505831 & -0.0148461575 & -0.0194874729 \\
        $i$159 & 0.12399451 & 0.1091634 & -0.073941205 & -0.154710841 & -0.0845474408 & 0.0090121649 \\
        $i$160 & 0.23591307 & 0.11228714 & -0.081897489 & -0.056886754 & -0.0432177179 & -0.1340466739 \\
        $i$161 & 0.15845492 & 0.06399078 & -0.080378105 & -0.035588615 & -0.0164507559 & -0.2280334333 \\
        $i$162 & 0.14051186 & 0.02945819 & -0.036437007 & -0.214928539 & 0.3897321725 & -0.1400835454 \\
        $i$163 & 0.5430053 & 0.07159714 & 0.006124041 & -0.239369256 & 0.4123487509 & -0.2892592607 \\
        $i$164 & 0.20169261 & 0.09786016 & -0.079619896 & -0.099657008 & 0.0100316874 & -0.0060493658 \\
        $i$165 & 0.15598481 & 0.12440081 & -0.076024864 & -0.114346156 & -0.0972785087 & 0.223420543 \\
        $i$166 & 0.12208871 & 0.07524293 & -0.072318932 & -0.121090722 & -0.0490116832 & 0.3115270691 \\
        $i$167 & 0.18920061 & 0.08040957 & -0.078791264 & -0.093892474 & -0.0232771606 & 0.0754791313 \\
        $i$168 & 0.45372708 & 0.076632 & -0.059065689 & -0.185273769 & 0.1339938814 & -0.1005946349 \\
        $i$169 & 0.21973708 & 0.13867545 & -0.075982408 & -0.049627388 & -0.0439245017 & 0.1301201449 \\
        $i$170 & 0.1183822 & 0.07988999 & -0.073811453 & -0.074450937 & -0.0944692762 & 0.3062850656 \\
        \hline
        \end{tabular}
\end{table*}
\begin{table*}
    \centering
    \caption{Final atmospheric parameters and radial velocities of all stars in the study sample. We performed a weighted mean to calculate the final values for stars with two or more spectra. Column 1 lists the identifiers used for the stars. Columns 2 and 3 list the coordinates in J200 epoch. Columns 4, 5, 6 and 7 list the final mean weighted PCA-based atmospheric parameters with uncertainties. Columns 8 and 9 list the calculated radial velocities and uncertainties, respectively. Column 10 lists the S/N of the spectra obtained for each star. Column 11 lists the 2MASS $J$ and $H$ quality flags. Column 12 lists the spectral types from SIMBAD. The full version is available as supplementary material.}
    \resizebox{16cm}{!}{
        \begin{tabular}{lccccccccccc}
        \hline
        \hline
        ID & RA & DEC & T$_{\text{eff}}^{\text{PCA}}$ & $\sigma$T$_{\text{eff}}^{\text{PCA}}$ & [Fe/H]$^{\text{PCA}}$ & $\sigma$[Fe/H]$^{\text{PCA}}$ & RV & $\sigma$RV & S/N & J,H & SpT$^{\text{SIMBAD}}$ \Tstrut\\
        & J2000 & J2000 & (K) & (K) & (dex) & (dex) & (km/s) & (km/s) & & 2MASS flag & \\
        \hline
        HIP 439	&	00:05:24.2	&	-37:21:25.8	&	3924	&	120	&	0.27	&	0.31	&	41.4	&	1.4	&	170	&	A,E	&	M2V	\\
        HIP 1242	&	00:15:28.0	&	-16:08:00.9	&	3341	&	224	&	-0.14	&	0.49	&	-24.7	&	1.7	&	210	&	A,A	&	M4V	\\
        HIP 1696	&	00:21:19.6	&	-45:44:46.6	&	3594	&	92	&	0.07	&	0.20	&	2.1	&	0.8	&	130	&	A,A	&	M1V	\\
        HIP 1842	&	00:23:18.5	&	-50:53:38.0	&	3981	&	173	&	0.35	&	0.29	&	-17.7	&	1.5	&	150	&	A,A	&	M4V	\\
        HIP 3261	&	00:41:30.5	&	-33:37:31.8	&	3250	&	89	&	-0.05	&	0.17	&	41.3	&	1.2	&	240	&	A,A	&	K9V	\\
        HIP 5496	&	01:10:22.8	&	-67:26:42.6	&	3480	&	93	&	-0.11	&	0.17	&	36.4	&	0.6	&	130,230	&	A,A	&	M2.5V	\\
        HIP 6351	&	01:21:34.5	&	-41:39:22.7	&	3280	&	87	&	-0.19	&	0.17	&	22.9	&	2.0	&	250	&	A,A	&	M0Vk	\\
        HIP 9724	&	02:05:04.9	&	-17:36:52.9	&	3421	&	94	&	-0.13	&	0.15	&	30.9	&	0.4	&	130,310	&	A,A	&	M2.5	\\
        HIP 9786	&	02:05:48.6	&	-30:10:36.1	&	3826	&	78	&	0.15	&	0.19	&	41.6	&	0.8	&	70,80,260	&	A,A	&	M2.5+V	\\
        L 225-057	&	02:34:21.2	&	-53:05:36.7	&	2909	&	212	&	0.17	&	0.39	&	41.2	&	0.9	&	190	&	A,A	&	M4	\\
        HIP 11964	&	02:34:22.6	&	-43:47:46.8	&	3925	&	93	&	0.50	&	0.17	&	0.2	&	2.3	&	380,330	&	A,A	&	K7V	\\
        GJ 3193	&	03:01:51.4	&	-16:35:35.7	&	3747	&	165	&	0.68	&	0.37	&	-6.1	&	1.0	&	230	&	A,A	&	M3	\\
        HIP 14555	&	03:07:55.8	&	-28:13:10.9	&	3112	&	98	&	-0.43	&	0.18	&	10.2	&	3.0	&	250	&	A,A	&	M1Ve	\\
        HIP 23452	&	05:02:28.5	&	-21:15:23.6	&	3163	&	87	&	0.10	&	0.16	&	-9.8	&	1.3	&	470	&	A,A	&	K7V	\\
        HIP 24186	&	05:11:40.5	&	-45:01:05.1	&	3360	&	88	&	-0.07	&	0.18	&	246.1	&	2.0	&	150	&	A,A	&	M1VIp	\\
        GJ 3357	&	05:36:00.1	&	-7:38:58.1	&	3722	&	204	&	0.51	&	0.36	&	27.8	&	1.1	&	160	&	A,A	&	M4	\\
        HIP 29295	&	06:10:34.6	&	-21:51:52.2	&	3433	&	119	&	0.19	&	0.46	&	13.7	&	0.8	&	170,570	&	E,D	&	M1V	\\
        HIP 30920	&	06:29:23.4	&	-2:48:50.0	&	3288	&	214	&	-0.34	&	0.44	&	31.6	&	1.9	&	160	&	A,A	&	M4.5V	\\
        HIP 31293	&	06:33:43.4	&	-75:37:48.3	&	3897	&	95	&	-0.12	&	0.18	&	11.7	&	0.6	&	70,310	&	A,A	&	M2V	\\
        HIP 31292	&	06:33:46.9	&	-75:37:30.2	&	3643	&	180	&	-0.36	&	0.48	&	10.1	&	0.71	&	70	&	A,A	&	M3V	\\
        HIP 31635	&	06:37:10.9	&	17:33:52.7	&	3768	&	89	&	0.37	&	0.18	&	-52.9	&	1.5	&	120	&	A,A	&	M0Ve	\\
        HIP 31862	&	06:39:37.6	&	-55:36:35.0	&	3919	&	89	&	-0.02	&	0.17	&	9.8	&	1.3	&	210	&	A,A	&	M0Vk	\\
        HIP 33499	&	06:57:46.6	&	-44:17:28.2	&	3401	&	131	&	-0.62	&	0.22	&	-21.5	&	0.7	&	340	&	A,A	&	M3.5V	\\
        HIP 36349	&	07:28:51.4	&	-30:14:49.1	&	3180	&	126	&	-0.09	&	0.25	&	31.1	&	2.4	&	240	&	A,A	&	M1Ve	\\
        HIP 37766	&	07:44:40.2	&	03:33:09.0	&	3528	&	193	&	0.01	&	0.30	&	30.2	&	2.3	&	300	&	A,A	&	M4Ve	\\
        GJ 300	&	08:12:40.9	&	-21:33:05.7	&	4031	&	201	&	0.31	&	0.30	&	2.6	&	1.3	&	170	&	A,A	&	M3.5V	\\
        HIP 40239	&	08:13:08.5	&	-13:55:00.6	&	3382	&	91	&	-0.09	&	0.19	&	8.3	&	1.7	&	190	&	A,A	&	M0V	\\
        GJ 3500	&	08:27:11.8	&	-44:59:21.5	&	3477	&	168	&	-0.06	&	0.25	&	50.4	&	3.6	&	190	&	A,A	&	M3	\\
        HIP 44722	&	09:06:45.4	&	-8:48:24.8	&	3181	&	92	&	0.02	&	0.19	&	40.4	&	1.8	&	250	&	A,A	&	K7V	\\
        HIP 45908	&	09:21:37.6	&	-60:16:55.1	&	2901	&	89	&	0.11	&	0.17	&	38.3	&	0.9	&	230	&	A,A	&	M0	\\
        HIP 46706	&	09:31:19.4	&	-13:29:19.3	&	3437	&	112	&	-0.46	&	0.17	&	13.8	&	1.2	&	330	&	A,A	&	M3V	\\
        HIP 47425	&	09:39:46.3	&	-41:04:03.0	&	3392	&	136	&	0.03	&	0.29	&	17.3	&	0.7	&	130	&	A,A	&	M3V	\\
        HIP 47780	&	09:44:29.9	&	-45:46:35.1	&	3274	&	117	&	0.00	&	0.26	&	47.5	&	0.9	&	170	&	A,A	&	M1	\\
        HIP 48336	&	09:51:09.6	&	-12:19:47.9	&	3608	&	90	&	-0.08	&	0.17	&	63.0	&	2.2	&	210	&	A,A	&	M1	\\
        HIP 52296	&	10:41:09.3	&	-36:53:43.6	&	3826	&	89	&	0.13	&	0.18	&	62.2	&	1.0	&	200	&	A,A	&	M0.5	\\
        HIP 52596	&	10:45:16.7	&	-30:48:26.9	&	3758	&	88	&	0.10	&	0.19	&	29.9	&	1.3	&	180	&	A,A	&	M1.5V	\\
        HIP 53767	&	11:00:04.3	&	22:49:59.3	&	3385	&	118	&	-0.18	&	0.25	&	16.1	&	0.5	&	180	&	A,A	&	M4V	\\
        GJ 410	&	11:02:38.3	&	21:58:01.7	&	3667	&	88	&	0.73	&	0.17	&	-8.9	&	1.9	&	310	&	A,A	&	M1Ve	\\
        HIP 54373	&	11:07:27.7	&	-19:17:29.4	&	3781	&	95	&	0.00	&	0.23	&	3.1	&	2.0	&	200	&	A,A	&	K5V	\\
        NLTT 26385	&	11:08:06.5	&	-5:13:46.9	&	3590	&	102	&	-0.09	&	0.19	&	25.7	&	1.2	&	170,180	&	A,A	&	M3V	\\
        HIP 55066	&	11:16:22.2	&	-14:41:35.9	&	3212	&	91	&	-0.26	&	0.18	&	17.8	&	1.6	&	140	&	A,A	&	K7V	\\
        HIP 55119	&	11:17:07.5	&	-27:48:48.5	&	3801	&	91	&	0.02	&	0.17	&	-22.0	&	1.5	&	210	&	A,A	&	K7	\\
        HIP 56244	&	11:31:46.6	&	-41:02:47.3	&	3501	&	156	&	-0.31	&	0.31	&	0.9	&	1.3	&	250	&	A,A	&	M3.5	\\
        HIP 56528	&	11:35:27.0	&	-32:32:23.3	&	3555	&	110	&	-0.68	&	0.25	&	18.2	&	0.9	&	170	&	A,A	&	M2V	\\
        HD 103932	&	11:57:56.2	&	-27:42:25.1	&	3663	&	64	&	0.00	&	0.25	&	49.5	&	0.6	&	220,250,320,300,280,530,170	&	E,D	&	K4+V	\\
        HIP 60559	&	12:24:52.4	&	-18:14:30.4	&	3335	&	121	&	0.04	&	0.30	&	51.0	&	0.9	&	70	&	A,A	&	M2	\\
        NLTT 31084	&	12:33:33.1	&	-48:26:11.2	&	3720	&	206	&	-0.07	&	0.45	&	6.1	&	2.1	&	220	&	A,A	&	M3.5	\\
        HIP 61629	&	12:37:52.3	&	-52:00:05.5	&	3297	&	98	&	0.44	&	0.21	&	-7.7	&	0.5	&	90,170	&	A,A	&	M3Ve	\\
        HIP 61874	&	12:40:46.3	&	-43:33:59.5	&	3324	&	113	&	-0.01	&	0.32	&	-34.6	&	0.7	&	80,90	&	A,A	&	M3V	\\
        HIP 62452	&	12:47:56.6	&	09:45:05.0	&	3617	&	196	&	0.66	&	0.48	&	22.2	&	1.0	&	110	&	A,A	&	M3.5Ve	\\
        WT 392	&	13:13:09.4	&	-41:30:39.7	&	3795	&	217	&	0.03	&	0.37	&	-35.5	&	2.1	&	180	&	A,A	&	M5	\\
        HIP 66675	&	13:40:07.2	&	-4:11:10.2	&	3640	&	95	&	0.08	&	0.23	&	23.4	&	1.6	&	210	&	A,A	&	K5V	\\
        HIP 66906	&	13:42:43.3	&	33:17:25.5	&	3565	&	160	&	-0.17	&	0.27	&	8.4	&	2.8	&	180	&	A,A	&	M4V	\\
        HIP 67164	&	13:45:50.7	&	-17:58:04.8	&	3884	&	146	&	0.29	&	0.22	&	5.0	&	0.9	&	190	&	A,A	&	M3.5	\\
        HD 120476a	&	13:49:04.0	&	26:58:47.0	&	3583	&	68	&	-0.13	&	0.15	&	-16.5	&	1.0	&	280,150	&	E,A	&	K4V	\\
        LTT 5381	&	13:51:23.2	&	-27:33:52.0	&	3650	&	80	&	-0.11	&	0.16	&	31.5	&	1.7	&	130,150	&	A,A	&	M3V	\\
        HIP 67960	&	13:55:02.6	&	-29:05:25.7	&	3761	&	92	&	-0.22	&	0.17	&	9.6	&	1.3	&	230	&	A,A	&	M0Vk	\\
        GJ 3820	&	13:59:10.5	&	-19:50:03.5	&	3476	&	200	&	-0.05	&	0.35	&	-15.0	&	2.8	&	170	&	A,A	&	M4.5V	\\
        HIP 68469	&	14:01:03.2	&	-2:39:18.1	&	3379	&	91	&	-0.23	&	0.18	&	-32.0	&	1.1	&	210	&	A,A	&	M0V	\\
        HIP 69454	&	14:13:12.9	&	-56:44:31.4	&	3431	&	74	&	-0.14	&	0.17	&	56.3	&	1.0	&	230,230	&	A,A	&	M2V	\\
        HIP 70890	&	14:29:42.9	&	-62:40:46.5	&	3498	&	254	&	-0.01	&	0.57	&	-24.2	&	1.3	&	180,190	&	A,A	&	M5.5Ve	\\
        HIP 70956	&	14:30:47.8	&	-8:38:46.6	&	3739	&	93	&	-0.08	&	0.19	&	-22.1	&	1.5	&	290	&	A,A	&	K7V	\\
        HIP 70975	&	14:31:01.2	&	-12:17:45.2	&	3947	&	168	&	0.49	&	0.30	&	0.6	&	1.2	&	210	&	A,A	&	M4	\\
        HIP 72509	&	14:49:31.6	&	-26:06:32.2	&	3798	&	99	&	-0.31	&	0.19	&	-35.5	&	2.8	&	150	&	A,A	&	M1.5V	\\
        HIP 72511	&	14:49:33.4	&	-26:06:20.5	&	3353	&	66	&	0.16	&	0.14	&	-31.8	&	1.5	&	160,130	&	A,A	&	M1.5V	\\
        HIP 72944	&	14:54:29.2	&	16:06:04.0	&	3223	&	111	&	-0.12	&	0.25	&	5.1	&	0.9	&	160	&	A,A	&	M3V	\\
        HIP 73182	&	14:57:26.4	&	-21:24:38.4	&	3391	&	125	&	-0.01	&	0.49	&	29.6	&	1.2	&	470,160	&	D,C	&	M1.5V	\\
        HD 131977	&	14:57:27.9	&	-21:24:52.6	&	3684	&	87	&	0.24	&	0.33	&	30.6	&	0.9	&	450,540,190,260	&	D,C	&	K4V	\\
        LP 682-018	&	15:15:43.7	&	-7:25:20.8	&	3889	&	195	&	-0.10	&	0.33	&	-15.4	&	1.8	&	160	&	A,A	&	M4V	\\
        GJ 3900	&	15:19:11.8	&	-12:45:06.2	&	3506	&	267	&	-0.02	&	0.63	&	-27.0	&	2.8	&	150	&	A,A	&	M4	\\
        HIP 76074	&	15:32:13.0	&	-41:16:31.5	&	3274	&	119	&	0.20	&	0.25	&	33.9	&	0.6	&	200	&	A,A	&	M2.5V	\\
        GJ 592	&	15:36:58.7	&	-14:08:00.6	&	3373	&	199	&	-0.47	&	0.37	&	1.8	&	1.1	&	180	&	A,A	&	M4	\\
        HIP 76901	&	15:42:06.8	&	-19:28:16.7	&	3964	&	126	&	0.22	&	0.23	&	84.7	&	1.0	&	150,140	&	A,A	&	M3	\\
        HIP 78353	&	15:59:53.4	&	-8:15:11.4	&	3802	&	91	&	-0.01	&	0.2	&	-12.1	&	1.3	&	190	&	A,A	&	M1V	\\
        \hline
        \end{tabular}
    }
\label{tab:literature_study_sample}
\end{table*}

\begin{table*}
    \centering
    \resizebox{16cm}{!}{
        \begin{tabular}{lccccccccccc}
        \hline
        \hline
        ID & RA & DEC & T$_{\text{eff}}^{\text{PCA}}$ & $\sigma$T$_{\text{eff}}^{\text{PCA}}$ & [Fe/H]$^{\text{PCA}}$ & $\sigma$[Fe/H]$^{\text{PCA}}$ & RV & $\sigma$RV & S/N & J,H & SpT$^{\text{SIMBAD}}$ \Tstrut\\
        & J2000 & J2000 & (K) & (K) & (dex) & (dex) & (km/s) & (km/s) & & 2MASS flag & \\
        \hline
                HIP 79431	&	16:12:41.8	&	-18:52:31.7	&	3812	&	166	&	0.02	&	0.31	&	0.9	&	1.2	&	250	&	A,A	&	M3V	\\
        GJ 618b	&	16:20:03.2	&	-37:31:48.6	&	3547	&	246	&	-0.25	&	0.55	&	33.9	&	1.1	&	20	&	A,A	&	M5	\\
        GJ 618a	&	16:20:03.5	&	-37:31:45.0	&	3715	&	66	&	-0.19	&	0.11	&	31.5	&	0.3	&	260,230,270,110	&	A,A	&	M3V	\\
        HIP 80268	&	16:23:07.7	&	-24:42:34.0	&	3963	&	118	&	0.40	&	0.26	&	-31.4	&	2.0	&	160	&	A,A	&	M2V	\\
        HIP 80440	&	16:25:13.0	&	-21:56:14.2	&	3461	&	91	&	0.21	&	0.18	&	-57.8	&	0.9	&	100	&	A,A	&	M0V	\\
        HIP 82283	&	16:48:46.0	&	-15:44:19.9	&	3458	&	89	&	-0.16	&	0.17	&	-73.5	&	1.6	&	160	&	A,A	&	M1.5Vk	\\
        HIP 82817	&	16:55:28.8	&	-8:20:10.3	&	2813	&	100	&	1.64	&	0.17	&	8.4	&	0.77	&	310,190	&	E,E	&	M3.5Ve	\\
        HIP 82834	&	16:55:38.0	&	-32:04:03.3	&	3601	&	91	&	-0.03	&	0.17	&	-22.0	&	1.6	&	200	&	A,A	&	K8Vk	\\
        HIP 83599	&	17:05:13.8	&	-5:05:38.6	&	3356	&	66	&	-0.08	&	0.13	&	38.2	&	1.1	&	300,280	&	A,A	&	M1.5V	\\
        HIP 84051	&	17:10:59.2	&	-52:30:56.0	&	3356	&	88	&	0.26	&	0.17	&	10.7	&	1.2	&	140	&	A,A	&	M1V	\\
        HIP 84123	&	17:11:52.3	&	-1:51:05.7	&	3270	&	105	&	0.24	&	0.19	&	-107.7	&	2.6	&	110	&	A,A	&	M3V	\\
        NLTT 44470	&	17:16:20.6	&	-5:23:51.3	&	2831	&	128	&	0.09	&	0.25	&	-36.5	&	1.1	&	190	&	A,A	&	M4V	\\
        GJ 3999	&	17:17:45.3	&	-11:48:54.2	&	3562	&	546	&	-0.05	&	0.99	&	22.1	&	1.5	&	300	&	A,A	&	M3	\\
        HIP 85523	&	17:28:39.9	&	-46:53:42.5	&	3488	&	109	&	0.17	&	0.26	&	-0.1	&	0.6	&	210,150	&	A,A	&	M3V	\\
        HIP 85605	&	17:29:36.3	&	24:39:11.1	&	3261	&	97	&	-0.43	&	0.20	&	-18.9	&	3.4	&	170	&	A,A	&	-	\\
        HIP 85647	&	17:30:11.2	&	-51:38:13.2	&	3844	&	92	&	0.10	&	0.17	&	-43.9	&	1.5	&	200	&	A,A	&	M0V	\\
        HIP 86057	&	17:35:13.6	&	-48:40:51.0	&	3721	&	103	&	-0.15	&	0.23	&	-17.1	&	0.7	&	120	&	A,A	&	M3V	\\
        HIP 86214	&	17:37:03.7	&	-44:19:08.8	&	3903	&	225	&	0.09	&	0.54	&	-43.3	&	1.4	&	120	&	A,A	&	M3.5	\\
        HIP 86961	&	17:46:12.8	&	-32:06:08.9	&	3702	&	76	&	0.28	&	0.16	&	-15.9	&	1.2	&	230,130	&	A,A	&	M1.5	\\
        HIP 86963	&	17:46:14.4	&	-32:06:08.2	&	3642	&	86	&	0.01	&	0.15	&	-22.7	&	0.7	&	180,130,230	&	A,A	&	M2.5	\\
        LTT 7077	&	17:46:29.3	&	-8:42:36.2	&	3175	&	188	&	-0.29	&	0.38	&	-16.6	&	2.0	&	190	&	A,A	&	M3.5	\\
        CD 57-6997	&	17:46:29.3	&	-57:19:24.3	&	3922	&	90	&	0.38	&	0.18	&	57.7	&	2.3	&	170	&	A,A	&	-	\\
        HIP 86990	&	17:46:34.3	&	-57:19:08.1	&	4011	&	105	&	0.45	&	0.20	&	-45.1	&	0.4	&	130,170	&	A,A	&	M3.5V	\\
        LTT 7246	&	18:15:12.4	&	-19:24:06.4	&	3242	&	112	&	-0.12	&	0.22	&	34.0	&	1.3	&	280	&	A,A	&	M2	\\
        GJ 714	&	18:30:12.0	&	-58:16:27.5	&	3287	&	93	&	0.20	&	0.21	&	-7.2	&	1.2	&	190	&	A,A	&	M1V	\\
        HIP 91608	&	18:40:57.3	&	-13:22:45.6	&	3821	&	92	&	0.08	&	0.20	&	-33.8	&	1.5	&	230	&	A,A	&	M1	\\
        LHS 5341	&	18:43:07.0	&	-54:36:48.2	&	3415	&	186	&	-0.49	&	0.35	&	16.6	&	1.0	&	170	&	A,A	&	M5	\\
        LTT 7419	&	18:43:12.5	&	-33:22:46.3	&	4010	&	88	&	0.45	&	0.17	&	-10.8	&	1.7	&	220	&	A,A	&	M2	\\
        GJ 4074	&	18:45:57.5	&	-28:55:53.3	&	3834	&	228	&	0.11	&	0.53	&	-25.7	&	2.3	&	190	&	A,A	&	M4	\\
        HD 175224	&	18:57:30.6	&	-55:59:30.2	&	3458	&	66	&	-0.18	&	0.16	&	-20.0	&	0.7	&	200,160	&	A,A	&	K5Ve+K7Ve	\\
        HIP 93206	&	18:59:07.5	&	-48:16:28.0	&	3935	&	172	&	0.05	&	0.33	&	-5.0	&	1.5	&	260	&	A,A	&	M4V	\\
        HIP 94739	&	19:16:42.9	&	-45:53:21.4	&	3715	&	89	&	-0.14	&	0.17	&	-37.8	&	1.7	&	240	&	A,A	&	M0Vk	\\
        GJ 754	&	19:20:48.0	&	-45:33:28.3	&	3531	&	220	&	-0.12	&	0.29	&	3.2	&	2.4	&	200	&	A,A	&	M4.5	\\
        HIP 99701	&	20:13:53.4	&	-45:09:50.6	&	3364	&	113	&	0.04	&	0.44	&	-26.0	&	1.0	&	150,690	&	A,D	&	M0V	\\
        HIP 100490	&	20:22:41.9	&	-58:17:08.4	&	3728	&	87	&	-0.08	&	0.16	&	1.8	&	1.9	&	220	&	A,A	&	M1V	\\
        L 210-70	&	20:27:42.1	&	-56:27:26.2	&	3540	&	220	&	0.04	&	0.41	&	-22.7	&	2.4	&	180	&	A,A	&	M2.5	\\
        HIP 102235	&	20:42:57.1	&	-18:55:04.8	&	3473	&	92	&	0.43	&	0.17	&	1.7	&	1.8	&	220	&	A,A	&	M1.5	\\
        HIP 102409	&	20:45:09.5	&	-31:20:26.7	&	3569	&	50	&	-0.10	&	0.10	&	-2.0	&	1.0	&	340,530,270,170,310	&	A,A	&	M1VeBa1	\\
        GJ 810a	&	20:55:37.7	&	-14:02:07.8	&	3188	&	241	&	0.06	&	0.49	&	-136.8	&	2.0	&	190	&	A,A	&	M4V	\\
        HIP 105090	&	21:17:15.3	&	-38:52:02.2	&	3410	&	80	&	0.01	&	0.31	&	20.5	&	0.3	&	380,290,430,660,220	&	D,C	&	M1V	\\
        HIP 106106	&	21:29:36.7	&	17:38:35.4	&	3531	&	163	&	-0.20	&	0.33	&	-20.0	&	0.7	&	150	&	A,A	&	M3Ve	\\
        HIP 106255	&	21:31:18.6	&	-9:47:26.3	&	3149	&	246	&	-0.33	&	0.50	&	-51.2	&	2.2	&	130	&	A,A	&	M4.5V	\\
        HIP 106440	&	21:33:34.0	&	-49:00:32.4	&	3467	&	86	&	0.23	&	0.32	&	13.9	&	0.6	&	140,200,190,190,340	&	A,D	&	M2/3V	\\
        GJ 1263	&	21:46:40.4	&	00:10:23.4	&	3949	&	216	&	0.11	&	0.43	&	-23.4	&	1.6	&	180	&	A,A	&	M4	\\
        HIP 107705	&	21:49:05.9	&	-72:06:08.6	&	3742	&	90	&	0.60	&	0.16	&	-15.4	&	2.6	&	260	&	A,A	&	M1Ve	\\
        HIP 108569	&	21:59:34.8	&	-59:45:10.5	&	3657	&	98	&	0.05	&	0.21	&	11.3	&	1.8	&	190	&	A,A	&	M2V	\\
        HIP 108706	&	22:01:13.1	&	28:18:24.9	&	3341	&	192	&	-0.20	&	0.45	&	-3.5	&	3.3	&	110	&	A,A	&	M3.5Ve	\\
        GJ 4248	&	22:02:29.4	&	-37:04:51.2	&	3700	&	108	&	-0.13	&	0.23	&	-17.6	&	0.8	&	180,190	&	A,A	&	M3V	\\
        HIP 110534	&	22:23:33.3	&	-57:13:14.7	&	3364	&	62	&	-0.10	&	0.12	&	-2.7	&	0.9	&	100,230	&	A,A	&	M1Vk	\\
        GJ 9780	&	22:25:05.0	&	-47:52:46.2	&	3617	&	245	&	0.13	&	0.47	&	-9.9	&	2.2	&	170	&	A,A	&	M3.5	\\
        HIP 111766	&	22:38:29.7	&	-65:22:42.3	&	3348	&	131	&	-0.14	&	0.31	&	-7.0	&	1.4	&	160,180	&	A,A	&	M3Ve	\\
        GJ 866	&	22:38:33.7	&	-15:17:57.3	&	3480	&	305	&	0.16	&	0.46	&	6824.7	&	3.3	&	200	&	A,A	&	M5V	\\
        GJ 867b	&	22:38:45.3	&	-20:36:51.9	&	3536	&	111	&	-0.09	&	0.26	&	0.4	&	1.0	&	260,110	&	A,A	&	M3.5V	\\
        GJ 867a	&	22:38:45.6	&	-20:37:16.1	&	3487	&	64	&	-0.01	&	0.15	&	8.8	&	1.1	&	250,350,130	&	A,A	&	M0Vep	\\
        HIP 113229	&	22:55:45.3	&	-75:27:32.1	&	3832	&	166	&	0.13	&	0.35	&	65.1	&	0.9	&	170	&	A,A	&	M3V	\\
        HIP 113576	&	23:00:16.1	&	-22:31:27.6	&	3505	&	99	&	0.20	&	0.25	&	15.7	&	1.3	&	120	&	A,E	&	K7+Vk	\\
        HIP 114719	&	23:14:16.6	&	-19:38:39.4	&	3249	&	94	&	0.24	&	0.19	&	4.9	&	0.7	&	140	&	A,A	&	M0.5Vk	\\
        GJ 896b	&	23:31:52.4	&	19:56:13.8	&	3182	&	305	&	0.24	&	0.89	&	-1.3	&	3.1	&	40	&	A,A	&	M4Ve	\\
        HIP 117828	&	23:53:50.2	&	-75:37:57.5	&	3395	&	125	&	-0.39	&	0.25	&	-13.3	&	1.9	&	230	&	A,A	&	M3	\\
        HIP 117966	&	23:55:39.8	&	-6:08:32.8	&	3198	&	125	&	0.04	&	0.27	&	19.5	&	2.2	&	220	&	A,A	&	M2.5Vk	\\
        \hline
        \end{tabular}
    }
\end{table*}        

\bsp	
\label{lastpage}
\end{document}